\documentclass[11pt, a4paper]{article}
\usepackage[english]{babel}
\usepackage{tikz}
\usepackage[compat=1.1.0]{tikz-feynman}
\usepackage{caption}
\usepackage{jheppub}
\usepackage{a4wide}
\usepackage{float}
\usepackage{bbold}
\usepackage{graphicx}
\usepackage{amssymb}
\usepackage{adjustbox}
\usepackage{subcaption}
\usepackage[normalem]{ulem}

\usepackage{textcomp}
\usepackage{rotating}
\usepackage{amsmath}
\usepackage{verbatim}
\usepackage{tabu}
\usepackage{cleveref}
\usepackage{rotating}
\usepackage{physics}
\usepackage{slashed}
\usepackage{amsthm}
\usepackage{csquotes}
\usepackage{url}
\usepackage{braket}
\usepackage{xcolor}
\usepackage{gensymb}
\usepackage{placeins}

\newcommand{\stkout}[1]{\ifmmode\text{\sout{\ensuremath{#1}}}\else\sout{#1}\fi}
\usepackage{makecell}

\usepackage{enumitem}
\setlist{itemsep=2em}
\newcommand{\bdphiks}{B_d^0 \to \phi K_S}
\newcommand{\bsphiks}{B_s^0 \to \phi K_S}

\newcommand{\bplusphikplus}{B^+\to\phi K^+}

\title{\boldmath Probing Isospin-Breaking New CP-violating Physics in $B\to \phi K$ Decays}

\author[a,b]{Robert Fleischer}\emailAdd{robert.fleischer@nikhef.nl}
\author[a]{Jelle Groot}\emailAdd{j.groot4@uva.nl}
\author[a,c]{K. Keri Vos}\emailAdd{k.vos@nikhef.nl}
\affiliation[a]{Nikhef, Science Park 105, NL-1098 XG Amsterdam, Netherlands}
\affiliation[b]{Department of Physics and Astronomy, Vrije Universiteit Amsterdam,
NL-1081 HV Amsterdam, Netherlands}
\affiliation[c]{Gravitational Waves and Fundamental Physics (GWFP),
Maastricht University, Duboisdomein 30,
NL-6229 GT Maastricht, the Netherlands}

\begin{document}

\begin{titlepage}

\vspace*{-2.0truecm}

\begin{flushright}
Nikhef-2026-004 
\end{flushright}

\vspace*{1.3truecm}

\begin{center}
\Large \bf{
CP Violation in \boldmath$B_{(s)}\to\phi K$\unboldmath\ Decays: Standard Model Benchmarks and Isospin-Breaking New Physics
}
\end{center}

\vspace{0.9truecm}

\begin{center}
{\bf Robert Fleischer\,${}^{a,b}$, Jelle Groot\,${}^{a,c}$ and K. Keri Vos\,${}^{a,d}$ 

\vspace{0.5truecm}

${}^a${\sl Nikhef, Science Park 105, NL-1098 XG Amsterdam, Netherlands}

${}^b${\sl  Department of Physics and Astronomy, Vrije Universiteit Amsterdam,\\
NL-1081 HV Amsterdam, Netherlands}

${}^c${\sl Institute for Theoretical Physics Amsterdam and Delta Institute for Theoretical
	Physics, \\ University of Amsterdam, Science Park 904, 1098 XH Amsterdam, 
	Netherlands}

{\sl $^d$Gravitational 
Waves and Fundamental Physics (GWFP),\\ 
Maastricht University, Duboisdomein 30,\\ 
NL-6229 GT Maastricht, Netherlands}\\[0.3cm]

}\end{center}

\vspace*{1.7cm}

%%%%%%%%%%%%%%%%%%%%%%%%%%%%%%%%%%%%%%%%%%%%%%%%%%%%%%%%%%

\begin{abstract}
\noindent
The penguin loop-suppressed $B\to\phi K$ decays are highly sensitive to contributions of hypothetical heavy new particles. Particularly interesting probes for testing the Standard Model and revealing such phenomena are provided by CP violation in the $B^0_d\to\phi K_{\rm S}$ decay. Standard-Model estimates for the corresponding CP-violating observables are theoretically limited by doubly Cabibbo-suppressed penguin contributions. We study these effects using a factorization approach, and provide predictions for CP asymmetries, to be contrasted with future measurements. To gain additional insight into these hadronic effects, we propose the $B_s^0\to\phi K_{\rm S}$ decay as a new channel. We predict the observables for this decay for which currently no measurements exist. By comparing $B^0_d\to\phi K_{\rm S}$ and $B^+ \to \phi K^+$, we further derive state-of-the-art constraints on isospin observables within the Standard Model. The same framework enables probing of possible New-Physics contributions, including general effects and those with non-trivial isospin structure. Interesting prospects arise for the high-precision era of flavour physics ahead.
\end{abstract}

%%%%%%%%%%%%%%%%%%%%%%%%%%%%%%%%%%%%%%%%%%%%%%%%%%%%%%%%%%

\vspace*{2.1truecm}

\vfill

\noindent
March 2026

\end{titlepage}

\vbox{}

\thispagestyle{empty}

\setcounter{page}{0}

%%%%%%%%%%%%%%%%%%%%%%%%%%%%%%%%%%%%%%%%%%%%%%%%%%%%%%%%%%

\newpage
\newpage

\section{Introduction}
CP violation in non-leptonic $B$-meson decays offers interesting probes of the Standard Model (SM) and searches for indirect signals of New Physics (NP) (for a recent review, see Ref.~\cite{Fleischer:2024uru}). 
In this paper, we will focus on charged and neutral 
$B\to \phi K$ decays, which are governed by QCD penguin topologies and originate from $b\to s \bar s  s$ quark-level transitions.
Since these flavour-changing neutral-current processes arise at the loop level in the SM, they are particularly promising portals for NP to manifest itself \cite{Fleischer:1996bv, Grossman:1997gr,Grossman:2003qp, Fleischer:2001pc, Biswas:2024bhn} .

In the search for possible footprints of NP, precision is the key limiting factor. 
An important observable is offered by mixing-induced CP violation in the $B^0_d\to \phi K_{\rm S}$ decay, which is generated by interference between $B^0_d\to \phi K_{\rm }$ and $\bar B^0_d\to \phi K_{\rm }$ decay processes through $B^0_d$--$\bar B^0_d$ mixing. 
For SM predictions, the dominant source of theoretical uncertainty arises from doubly Cabibbo-suppressed penguin contributions, which are usually neglected.
Given the increasing experimental precision and excellent future prospects, it is important to examine such effects carefully. We estimate these effects in $B\to \phi K$ decays and calculate the corresponding observables in the SM, which serve as references for comparison with the measurements. 

To obtain additional insights into the hadronic penguin contributions to the $B\to\phi K$ decays, we propose the new $B^0_s\to \phi K_{\rm S}$ decay channel, which originates from $b\to d \bar s  s$ quark-level transitions. In contrast to the $B^0_d\to \phi K_{\rm S}$ mode, the hadronic penguin effects in $B^0_s\to \phi K_{\rm S}$ are not doubly Cabibbo-suppressed and therefore contribute more significantly to the total branching ratio and CP asymmetries, offering a clearer experimental handle on their size once measured. This situation is similar to the CP-violation measurements  $B^0_d \to J/\psi K_{\rm S}$ and $B^0_s\to J/\psi \phi$ decays, where corrections from doubly Cabibbo-suppressed penguin topologies can be included through strategies involving control channels and the use of the $SU(3)$ flavour symmetry (for a recent study, see Ref.~\cite{DeBruyn:2025rhk}). At the moment, there is no experimental information available for the $B^0_s\to \phi K_{\rm S}$ mode. Therefore, we highly encourage experimental studies of this decay.  

We complement the analysis of the neutral $B^0_d\to \phi K_{\rm S}$ decay with its charged counterpart $B^+\to\phi K^+$. The comparison of these modes probes the isospin structure of the underlying decay amplitudes. Following Ref.~\cite{Fleischer:2001pc}, we first perform a SM analysis and decompose the decay amplitudes into $I=0$ and $I=1$ components. We analyse a corresponding set of isospin-sensitive observables and obtain state-of-the-art constraints using current experimental data.

Subsequently, we use this framework to study possible NP contributions to these decays across different isospin configurations in a model-independent way. To this end, the NP decay amplitudes are likewise decomposed into components with definite isospin \cite{Fleischer:2001pc}. We consider two cases: NP contributions arising purely from the $I=0$ sector and NP purely from the $I=1$ sector. For these scenarios, we derive constraints on the NP parameters and the associated observables from current measurements, which still leave substantial room for NP to enter. Therefore, these decays offer an interesting opportunity to further explore in the upcoming high-precision era of flavour physics.

The outline of this paper is as follows:  In Section~\ref{sec:ReviewSM}, we discuss the theoretical framework of our analysis of the $B^0_{(s)}\to \phi K$ decays in the SM, while we give the corresponding numerical results in Section~\ref{sec:numerical}. In Section~\ref{sec:isospinSM}, we introduce the isospin-dependent observables combining the decay information of the $B_d^0\to\phi K_{\rm S}$ and $B^+\to\phi K^+$ decays. We perform the NP isospin analysis in Sec.~\ref{sec:NewPhysics}. Our conclusions are summarized in Sec.~\ref{sec:conclusions}.

\section{\texorpdfstring{\boldmath $B\to\phi K$ decays in the SM}{} }
\label{sec:ReviewSM} 
\begin{figure}[t!]
    \centering
    \begin{subfigure}{0.4\textwidth}
        \centering
        \includegraphics[width=\textwidth]{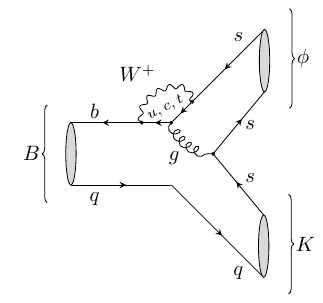}
        \caption{QCD penguin}
        \label{subfig:QCDpenguin}
    \end{subfigure}
    \begin{subfigure}{0.4\textwidth}
        \centering
        \includegraphics[width=\textwidth]{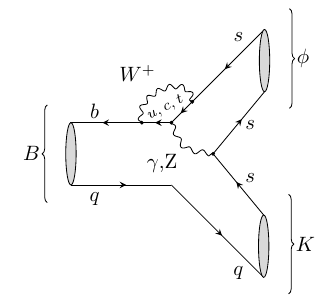}
        \caption{Colour-suppressed EW penguin}
        \label{subfig:ColSupEWpenguin}
    \end{subfigure}
    
    \begin{subfigure}{0.4\textwidth}
        \centering
        \includegraphics[width=\textwidth]{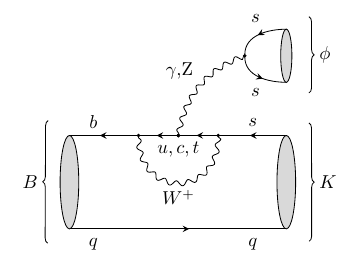}
        \caption{Colour-allowed EW penguin}
        \label{subfig:ColAllEWpenguin}
    \end{subfigure}
    \begin{subfigure}{0.4\textwidth}
        \centering
        \includegraphics[width=\textwidth]{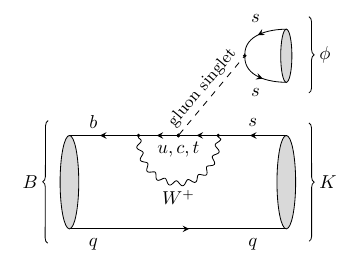}
        \caption{Colour singlet QCD penguin}
        \label{subfig:GluonSinglet}
    \end{subfigure}
    
    \caption{SM topologies contributing to $B\to\phi K$ processes via $b\to s$ quark transitions, where $B=B^+,B_d^0$ and $K=K^+,K^0$ for $q=u,d$, respectively.}
    \label{fig:PenguinDiagrams}
\end{figure}

In the SM, penguin topologies dominate $B\to \phi K$ decays. These include QCD penguins, colour-allowed electroweak penguins (EWP), colour-suppressed electroweak penguins, and colour-singlet QCD penguins. We first focus on $B\to\phi K$ processes that proceed through a $b\to s$ quark transition: the charged $B^+\to\phi K^+$ channel and the neutral $B_d^0\to \phi K_{\rm S}$ decay.\footnote{In the following, we treat the $\phi$ as a pure $s\bar{s}$ state and do not include the small $\omega-\phi$ mixing, see Ref.~\cite{Faller:2008gt}.} The decay topologies for these two processes differ only in the flavour of the spectator quark with $q=u$ for the charged mode and $q=d$ for the neutral mode, as shown in Fig.~\ref{fig:PenguinDiagrams}. 

Generically, utilizing the unitarity of the CKM matrix and not distinguishing the isospin states, the decay amplitudes of $B\to\phi K$ processes via a $b\to s$ quark transition can be parametrized as 
\begin{align}
\label{eq:decayamp}
    A(B\to\phi K)&=\frac{\mathcal{A}_P^\prime}{\sqrt{\epsilon}}\left[1+\epsilon b e^{i\theta}e^{i\gamma}\right] \ ,
\end{align}
where $\epsilon \equiv \lambda^2/(1-\lambda^2)=0.05$ is expressed in terms of the Wolfenstein CKM parameter $\lambda \equiv |V_{us}| = 0.223$ \cite{DeBruyn:2022zhw}, and $\gamma$ denotes the angle of the unitarity triangle (UT). We use \cite{LHCb:2021dcr}
\begin{equation}\label{eq:gamma}
    \gamma_{B\to D K}= (64.9 \pm 4.5)^\circ \ ,
\end{equation}
as determined from the pure tree-level $B\to D K$ decays through a time-independent analysis.
We have further defined 
\begin{align}
\label{eq:apprime}
    \mathcal{A}_P^{'}\equiv A\lambda^3 P^{(ct)'},
\end{align}
where $A\equiv |V_{cb}|/\lambda^2$. There are long-standing discrepancies between the exclusive and inclusive determinations of $|V_{ub}|$ and $|V_{cb}|$~\cite{Gambino:2019sif, Gambino:2020jvv, Bordone:2019vic}, which lead to two determinations for $A$; see Ref.~\cite{DeBruyn:2022zhw} for a detailed discussion. Using the inclusive value of $|V_{cb}|$ from Ref.~\cite{Bordone:2021oof} (see also Ref.~\cite{Finauri:2023kte})
and the exclusive value from HFLAV, we find
\begin{equation}\label{eq:Aincexcl}
       A|_{\rm incl} = 0.85 \pm 0.01\ , \quad \quad A|_{\rm excl} = 0.80 \pm 0.01 \ .
\end{equation}

The hadronic penguin parameters $b$ and CP-conserving strong phase $\theta$ in Eq.~\eqref{eq:decayamp} are doubly Cabibbo-suppressed by the factor $\epsilon$. They are defined through
\begin{align}
\label{eq:hadronicParameters}
        be^{i\theta}\equiv R_b\left[\frac{P^{(ut)'}}{P^{(ct)'}}\right]\ 
\end{align}
with the penguin parameters
\begin{align}\label{eq:pencoef}
P^{(qt)^\prime}\equiv P^{q^\prime}-P^{t^\prime},
\end{align}
where $P^q$ denotes a penguin transition with an internal $q$ quark, and the primes indicate a $b\to s$ quark transition to distinguish from $b \to d$ quark transitions discussed later. In the limit where the top-penguin contribution dominates, we have 
\begin{equation}
P^{(ut)\prime}=P^{(ct)\prime} \ , \quad\quad b= R_b\ . 
\end{equation}

In addition, we have
\begin{equation}
R_b\equiv \left(1-\frac{\lambda^2}{2}\right)\frac{1}{\lambda} \; \left|\frac{V_{ub}}{V_{cb}}\right| \ ,
\end{equation}
which denotes the side of the UT from the origin to the apex in the complex plane. For the inclusive and exclusive determination for $|V_{ub}|$ from Ref.~\cite{HeavyFlavorAveragingGroupHFLAV:2024ctg}, we find \begin{equation}\label{eq:hybrid}
    R_b|_{\rm incl} = 0.421 \pm 0.017\ , \quad \quad R_b|_{\rm excl} = 0.377 \pm 0.014 \ .
\end{equation}  
The tension between the inclusive and exclusive CKM factors, and the resulting different values of $A$, directly affects the branching ratio estimates. Accordingly, in the numerical analysis of the observables presented in this section, we will quote results for both the inclusive and the exclusive cases.

\subsection{Observables}
For neutral $B_q^0$ ($q=d,s$) mesons decaying into CP eigenstates $f$, the quantum-mechanical oscillations between the $B_q^0$ and $\bar{B}_q^0$ states give a time-dependent decay rate asymmetry \cite{Fleischer:2002ys}:
\begin{align} 
\label{eq:timeDepCPassym}
    \mathcal{A}_{\text{CP}}(t)&\equiv\frac{\Gamma(B_q^0(t)\to f) -\Gamma(\overline{B_q^0}(t)\to f)}{\Gamma(B_q^0(t)\to f) + \Gamma(\overline{B_q^0}(t)\to f)}\\&=\frac{\mathcal{A}^{\text{dir}}_\text{CP}(B_q^0\to f)\cos{\left(\Delta M_qt\right)}+\mathcal{A}^{\text{mix}}_\text{CP}(B_q^0\to f)\sin{\left(\Delta M_qt\right)}}{\cosh (\Delta\Gamma_q t/2) + \mathcal{A}^{\Delta\Gamma}_{\rm CP}(B_q^0\to f)\sinh (\Delta\Gamma_q t/2)}\:,
\end{align}
where $\Delta \Gamma_q\equiv \Gamma_L^{(q)} - \Gamma_H^{(q)}$ and  $\Delta M_q\equiv M_H^{(q)} - M_L^{(q)}$ describe the differences between the decay widths and masses of the heavy and light $B_q^0$-meson mass eigenstates. The direct and mixing-induced CP asymmetries are described by $\mathcal{A}^{\text{dir}}_\text{CP}$ and $\mathcal{A}^{\text{mix}}_\text{CP}$ respectively, and $\mathcal{A}^{\Delta\Gamma}_{\rm CP}$ is the decay-width-difference CP asymmetry. The decay-width difference in the $B_d^0$ system $y_d \equiv \Delta\Gamma_d/(2\Gamma_d) = \mathcal{O}(10^{-3})$ is negligibly small, but for the $B_s^0$ system $ y_s \equiv  \Delta\Gamma_s/(2 \Gamma_s) =0.062\pm0.004$ \cite{ParticleDataGroup:2024cfk}. The asymmetry $\mathcal{A}_{\Delta\Gamma}^{\rm CP}$ is therefore an interesting additional observable for $B_s^0$ decay modes.
We note that the three asymmetries are not independent from one another, satisfying the sum rule
\begin{equation}
 (\mathcal{A}_{\text{CP}}^{\text{dir}})^2+(\mathcal{A}_{\text{CP}}^{\text{dir}})^2 +   (\mathcal{A}^{\Delta\Gamma}_{\rm CP})^2 = 1\, .
\end{equation}  

The neutral final states $f$ considered in this work contain neutral kaons which are detected through their short- and long-lived weak eigenstates $K_{\rm S, L}$. We focus on the $K_{\rm S}$ component due to its shorter lifetime and higher reconstruction efficiency. We also neglect CP violation in the neutral kaon system and treat the $K_{\rm S}$ as CP-even. 

For the amplitude parameterization in Eq.~\eqref{eq:decayamp}, the CP asymmetries in the $B_d^0\to \phi K_{\rm S}$ system take the following form~\cite{Fleischer:1999zi}: 
\begin{align}
    \label{eq:AdirBdksphi}
\mathcal{A}_{\text{CP}}^{\text{dir}}(B_d^0\to \phi K_{\rm S})&=\frac{-2\epsilon b \sin \gamma \sin \theta}{1+2\epsilon b\cos\theta\cos\gamma+\epsilon^2b^2}\ , \\
\label{eq:AmixBdksphi}
\mathcal{A}_{\text{CP}}^{\text{mix}}(B_d^0\to \phi K_{\rm S})&= -\frac{\sin \phi_d+2\epsilon b \cos \theta \sin (\phi_d+\gamma)+\epsilon^2b^2\sin (\phi_d+2\gamma)}{1+2\epsilon b\cos\theta\cos \gamma + \epsilon^2 b^2}\ ,\\
\label{eq:ADeltaGammaBdksphi}
\mathcal{A}^{\Delta\Gamma}_{\rm CP}(B_d^0\to \phi K_{\rm S})&= \frac{\cos \phi_d+2\epsilon b \cos \theta \cos (\phi_d+\gamma)+\epsilon^2b^2\cos (\phi_d+2\gamma)}{1+2\epsilon b\cos\theta\cos \gamma + \epsilon^2 b^2}\ ,
\end{align}
where $\phi_d$ is the $B_d^0$--$\bar{B}_d^0$ mixing phase. This phase is determined from CP violation in the $B_d^0 \to J/\psi K_{\rm S}$ decay, including also doubly-Cabibbo suppressed penguin contributions \cite{Barel:2020jvf, Barel:2022wfr}, and $\phi_d$ is extracted from data as
\begin{equation}\label{eq:phid}
   \phi_d= (44.4^{+1.6}_{-1.5})^\circ. 
\end{equation}
The experimental values of the $\bdphiks$ and $B^+\to\phi K^+$ CP-violating observables and branching ratios are given in Table~\ref{tab:expvalues}.

\begin{table}[t!]
\small
\centering
\resizebox{0.7\textwidth}{!}{
{\tabulinesep=1.2mm
\begin{tabu}{c|cccc}
 Mode & $\mathcal{B}^{\rm SM}_{\rm exp}[10^{-6}]$ &$\mathcal{A}_{\text{CP}}^{\text{dir}}(B\to f)$ &   $\mathcal{A}_{\text{CP}}^{\text{mix}}(B_q^0\to f)$ \\
 \hline \hline
$B_d^0\to\phi K_{\rm S}$ & $3.7\pm0.4$ 
& $-0.09\pm0.12$ & $-0.58\pm0.12$  
\\
$B^+\to \phi K^+$ & $8.8^{+0.7}_{-0.6}$ & $-0.017\pm0.017$ & $-$\\
\end{tabu}}}
\caption{Experimental values for the branching ratios and CP-violating observables of the $\bdphiks$ and $B^+\to\phi K^+$ processes \cite{ParticleDataGroup:2024cfk}.  
}
\label{tab:expvalues}
\end{table}

It is useful to introduce an ``effective'' mixing phase $\phi^{\rm eff}_{d,\phi K_{\rm S}}$, that relates the direct and mixing-induced CP asymmetries through 
\begin{equation}
\label{eq:Defphidef}
    \sin\phi_{d, \phi K_{\rm S}}^{\rm eff} \equiv -\frac{\mathcal{A}_{\rm CP}^{\rm mix}(B_d\to \phi K_{\rm S})}{\sqrt{1-\left[\mathcal{A}_{\rm CP}^{\rm dir}(B_d\to \phi K_{\rm S})\right]^2}} \ . 
\end{equation}
The effective mixing phase is given by
 \begin{align}
 \label{eq:phideff}
\phi_{d,f}^{\text{eff}}\equiv\phi_d^{\rm SM} + \phi_d^{\rm NP} +\Delta \phi_d^f +\Delta \phi_d^{\rm NP}\ ,
 \end{align}
 where $\Delta \phi_d^f$ is a decay-channel specific non-perturbative hadronic phase shift:
\begin{align}
\label{eq:hadronTan}
    \tan\Delta\phi_d^{\phi K_{\rm S}}=\frac{2\epsilon b\cos\theta\sin\gamma +\epsilon^2b^2\sin2\gamma}{1+2\epsilon b\cos\theta\cos\gamma+\epsilon^2b^2\cos2\gamma} \ .
\end{align}
In addition, we parametrize
\begin{equation}
    \phi_d \equiv \phi_d^{\rm SM} + \phi_d^{\rm NP} = 2\beta + \phi_d^{\rm NP} \ ,
\end{equation}
where $\beta$ is one of the UT angles and $\phi_d^{\rm NP}$ probes possible CP-violating New Physics (NP) contributions in $B_d^0$--$\bar{B}_d^0$ mixing. Finally, $\Delta\phi_d^{\rm NP}$ probes NP contributions entering at the decay amplitude level. 

For the charged $B^+\to \phi K^+$ decay, only direct CP violation can occur. Consistent with Eq.~\eqref{eq:timeDepCPassym}, we define the direct CP asymmetry in this channel as\footnote{Our definition differs by a sign from the PDG convention~\cite{ParticleDataGroup:2024cfk}.}
\begin{align}
\label{eq:AdirBplus}
    \mathcal{A}^{\rm dir}_{\rm CP}(\bplusphikplus)=\frac{|A(\bplusphikplus)|^2-|A(B^-\to\phi K^-)|^2}{|A(\bplusphikplus)|^2+|A(B^-\to\phi K^-)|^2}\, .
\end{align}

The neutral and charged $B\to \phi K$ modes are related through isospin symmetry. For the dominant penguin modes, these decays only differ through their spectator $s$ versus $d$ quark, respectively. However, the charged $\bplusphikplus$ modes can also proceed through exchange and annihilation topologies, which are expected to be dynamically suppressed. Assuming the isospin-symmetry limit and neglecting annihilation modes, this implies 
\begin{equation}\label{eq:adirequal}
    \mathcal{A}_\text{CP}^{\text{dir}}(B^+\to\phi K^+)=\mathcal{A}_\text{CP}^{\text{dir}}(B_d^0\to\phi K_{\rm S}) \ ,
\end{equation}
which agrees with the experimental values in Table~\ref{tab:expvalues}. At the same time, the direct CP asymmetry of the neutral $\bdphiks$ mode still has a large uncertainty.

\subsection{Estimates of the hadronic matrix elements}\label{sec:fact}
The penguin matrix elements cannot reliably be computed from first principles, but their values can be estimated by factorizing the corresponding quark currents. Generally, the $B \to \phi K$ decay processes can be described by the low-energy effective  Hamiltonian~\cite{ Fleischer:1992gp,Fleischer:1993gr}
\begin{equation}\label{eq:Heff}
    \mathcal{H}_{\rm eff}(\Delta B=-1) = \frac{G_F}{\sqrt{2}} \sum_{q=u,c} \lambda_q \left\{C_1(\mu) \mathcal{O}_1^q + C_2(\mu) \mathcal{O}^q_2 + \sum_{k=3}^{10}C_k(\mu) \mathcal{O}_k\right\}\ ,
\end{equation}
where $\mu$ is the renormalization scale. For simplicity, do not explicitly write the $\mu$-dependence in the following. In addition, $\lambda_q \equiv V_{qs}V_{qb}^*$ are products of CKM matrix elements, $C_k$ are Wilson coefficients, and $\mathcal{O}_k$ are the corresponding SM four-quark operators. The operators $\mathcal{O}_k$ are defined in Ref.~\cite{Fleischer:1992gp,Fleischer:1993gr}, where $k=1,2$ are the current--current operators, while $k=3,4,5,6$ and $k=7,8,9,10$ are the QCD and electroweak penguin operators, respectively. 

\begin{figure}
    \centering
    \includegraphics[width=0.5\linewidth]{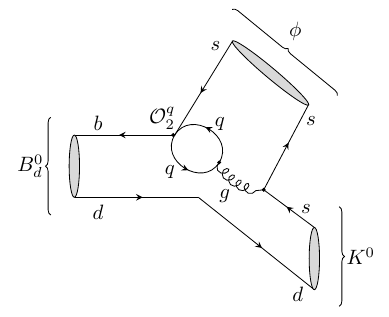}
    \caption{Illustration of penguin matrix elements of the $\mathcal{O}_2^q$ current--current operator contributions to the $B_d^0\to\phi K^0$ decay.}
    \label{fig:CCpenguin}
\end{figure}

To obtain a theoretical estimate of the $\bdphiks$ branching ratio, we express the decay amplitude in Eq.~\eqref{eq:decayamp} by writing~\cite{Fleischer:1993gr} 
\begin{align}\label{eq:Pctex}
    P^{(ct)'} =\frac{G_F}{\sqrt{2}}\mathcal{C}_{c,\text{SM}}\bra{\phi K^0}(\bar{s}s)_{\text{V-A}}(\bar{s}b)_{\text{V-A}}\ket{B_d^0}\:,
\end{align}
where we have defined
\begin{align}\label{eq:CSM}
   \mathcal{C}_{c,\text{SM}} &\equiv \frac{4}{3} \big(C_3 + C_4\big) + C_5 + \frac{1}{3}C_6 \nonumber \\& -\frac{1}{2}\left[C_7 +\frac{1}{3}C_8+\frac{4}{3}\big(C_9 + C_{10}\big)\right] + C_{\text{pen},c}^{\rm CC}  \ .
   \end{align}
Moreover, we introduced
\begin{equation}\label{eq:CCpen}
C_{\text{pen},q}^{\rm CC} \equiv  \frac{\alpha_s(m_b)}{8\pi} \left[\frac{10}{9} - G(m_q,k^2,m_b)\right] \left[\frac{8}{9}C_2 - \frac{28}{27} \frac{\alpha}{\alpha_s(m_b)} (3 C_1+ C_2)\right]  \end{equation}
with $q\in \{u,c\}$, which describes the contribution of the current--current operators at the loop level through penguin-like matrix elements. This process is depicted in Fig.~\ref{fig:CCpenguin}. We have defined the loop function 
   \begin{align}
   \label{eq:Gmu}
       G(m_q,k,\mu)\equiv-4\int_0^1 dx\, x (1-x)\ln\left[\frac{m_q^2-k^2x(1-x)}{\mu^2}\right],
   \end{align} where $k$ denotes the four-momentum of the virtual gluon (photon) in the QCD (EW) penguin topologies. 

To obtain a numerical estimate of the Wilson coefficients $\mathcal{C}_{c,\text{SM}}$ in Eq.~\eqref{eq:CSM}, we take the numerical SM values of $\Vec{C}\equiv \begin{pmatrix} C_{1},& \dots&, C_{10}\end{pmatrix}^T$ from Ref.~\cite{Buras:1998raa} in the NDR scheme at $\Lambda_{\overline{\rm MS}}^{(5)}=225$ MeV at $\mu=\overline{m}_b(m_b)=4.40\,$GeV. In the ``physical'' momentum range $1/4<k^2/m_b^2<1/2$~\cite{Bander:1979px, Gerard:1988jj}, we then find
    \begin{equation}
\label{eq:SMValuesWilsonVec2}
    |\mathcal{C}_{c,\text{SM}}| = [0.036,0.046]. \ %\jg{0.0415\pm0.0034 } .%0.041\pm0.005 \ ,
\end{equation}

The matrix element in Eq.~\eqref{eq:Pctex} can be factorized as
\begin{align}
\label{eq:factHadMatElem}
    \bra{\phi K^0}(\bar{s}s)_{\text{V-A}}(\bar{s}b)_{\text{V-A}}\ket{B_d^0} = a_{\rm NF}^0 \bra{\phi}(\bar{s}s)_{\text{V-A}}\ket{0}\bra{K^0}(\bar{s}b)_{\text{V-A}}\ket{B_d^0},
\end{align}
where 
 \begin{align}
\label{eq:fmepsilon}
    \bra{\phi(p_V)}\bar{s}\gamma^\mu s\ket{0}= f_\phi m_\phi \epsilon_\phi^\mu(p_V)
\end{align}
is expressed in terms of the $\phi$-meson decay constant $f_\phi= (228.5 \pm 3.6)\,$MeV \cite{ParticleDataGroup:2024cfk}, polarization vector $\epsilon_\phi$, and mass $m_\phi=1019.5\,$~MeV. The non-factorizable effects in the matrix elements are parametrized by
\begin{equation}\label{eq:anf}
 a_{\rm NF}^0 \equiv 1 + \delta_{\rm NF}^0   \ ,
\end{equation}
where the superscript $0$ refers to the charge of the kaon in the final state and $\delta_{\rm NF}^0$ parameterizes the magnitude of the non-factorizable contributions.
The hadronic $B\to K$ form factors can be written as 
\begin{align}
\label{eq:KsbB}
    \bra{K^0(p_P)}\bar{s}\gamma^\mu b\ket{B_d^0(p_B)}&=F^{B^0K^0}_1(q^2)\left[P^\mu-\frac{m_{B_d^0}^2-m_{K^0}^2}{q^2}q^\mu\right]
 + F^{B^0K^0}_0(q^2)\left[\frac{m_{B_d^0}^2-m_{K^0}^2}{q^2}\right]q^\mu \:,
\end{align}
where $P\equiv p_B + p_V$ and $q\equiv p_B - p_P=p_V$. Combining the above, we write the penguin parameter $P^{(ct)\prime}$ given in Eq.~\eqref{eq:Pctex} as 
\begin{equation}\label{eq:pct}
   P^{(ct)^\prime} = \frac{G_F}{\sqrt{2}} f_\phi m_\phi  \;  F_1^{B^0K^0}\!\left(m_\phi^2\right)\mathcal{C}_{c,\text{SM}}\; a_{\rm NF}^0 \;(P\cdot\epsilon_\phi) \ .
\end{equation}
Using Eq.~\eqref{eq:decayamp}, the CP-averaged branching ratio
\begin{align}
\mathcal{B}(B_q\to f)\equiv
\frac{\tau_{B_q}}{2}\left[\Gamma(B_q\to f)+\Gamma(\bar B_q\to f)\right]
\end{align}
yields for the $B_d^0\to \phi K_{\rm S}$ decay
\begin{align}
\label{eq:BRstandardModel}
\mathcal{B}\left(B_d^0\to \phi K_{\rm S}\right)
= \frac{\hat{C}^0}{2\epsilon}
\left[1+2\epsilon b\cos\theta\cos\gamma+(\epsilon b)^2\right].
\end{align}
Further using Eq.~\eqref{eq:pct}, we find 
\begin{align}
\label{eq:Chatdef}
    \hat{\mathcal{C}}^0 &= \tau_{B_d^0}\frac{G_F^2}{32\pi} A^2\lambda^6 m_{B_d^0}^3f_\phi^2\left[\Phi\left(m_{K^0}/m_{B_d^0}, m_\phi/m_{B_d^0}\right)\right]^3  \left[F^{B^0K^0}_1(m_\phi^2)\right]^2\left|\mathcal{C}_{c,\text{SM}} \right|^2 |a_{\rm NF}^0|^2 ,
\end{align}
where 
\begin{align}
    \Phi\left(X,Y\right) \equiv \sqrt{\left[1-\left(X+Y\right)^2\right]\left[1-\left(X-Y\right)^2\right]}
\end{align}
is the standard phase-space factor. 

For the charged $B^+\to \phi K^+$ branching ratio, the above derivation also holds, and a similar expression can be obtained by straightforward replacements introducing a parameter $a_{\rm NF}^+$ for the non-factorizable contributions.

\subsection{\texorpdfstring{\boldmath The $B_s^0\to\phi K_{\rm S}$ decay}{}}
Another interesting decay, also dominated by QCD penguins in the SM, is the $B_s^0\to \phi K_{\rm S}$ mode. This decay is similar to the $B_d^0\to \phi K_{\rm S}$ decay, but governed by a $b\to d$ quark transition and, consequently, different CKM elements enter. The decay amplitude of this process can be written as
\begin{align}
    A(B_s^0\to\phi K_{\rm S})=-\tilde{\mathcal{A}}_{P}^{(ct)}\left[1-\tilde{b}e^{i\tilde{\theta}}e^{i\gamma}\right],    
\end{align}
where the hadronic parameters $\tilde{\mathcal{A}}_{P}^{(ct)}$ and $ \tilde{b}e^{i\tilde{\theta}}$ are defined in analogy to Eqs.~\eqref{eq:apprime} and \eqref{eq:hadronicParameters}. 

The expressions for the CP-violating observables of this decay are obtained from those of the $B_d^0\to \phi K_{\rm S}$ channel in Eqs.~\eqref{eq:AdirBdksphi} and \eqref{eq:AmixBdksphi} by replacing
\begin{equation}
\epsilon \to -1 ,\qquad \phi_d\to \phi_s, \qquad b \to \tilde{b}, \qquad\theta \to \tilde{\theta} \ . 
\end{equation} 
For the $B_s^0-\bar{B}_s^0$ mixing phase, we use \cite{Barel:2022wfr}
\begin{equation}
    \phi_s = -(2.01\pm 0.12)^\circ\ ,  
\end{equation}
including penguin contributions.
In contrast to the $\bdphiks$ decay, the hadronic parameters in the $B_s^0\to\phi K_{\rm S}$ mode are not doubly Cabibbo-suppressed. This enhancement makes the $B_s^0$ channel particularly well-suited for studying these parameters.

Measurements of the direct and mixing-induced CP asymmetries of $B_s^0\to \phi K_{\rm S}$ could be used to directly determine $\tilde{b}$ and $\tilde\theta$ using $\gamma$ and $\phi_s$ as inputs. Unfortunately, the $B_s^0\to\phi K_{\rm S}$ decay has not yet been observed. 
We therefore propose measuring this mode and its CP asymmetries, as this would provide interesting insights into penguin parameters. In anticipation of these measurements, we predict its branching ratio in the SM, again assuming factorization of the leading $P^{(ct)}$ amplitude. 

\begin{figure}[t!]
    \centering
    \includegraphics[width=0.495\linewidth]{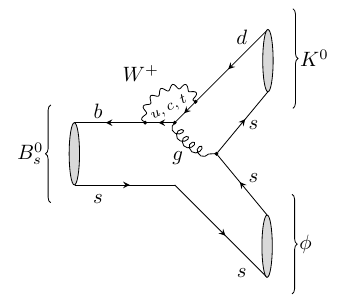}\includegraphics[width=0.495\linewidth]{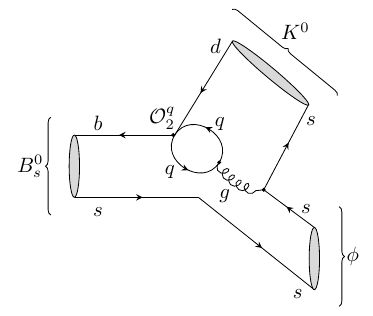}
    \caption{Contributions to the $B_s^0 \to \phi K_S$ decay in which the $B_s^0$ meson hadronizes into a $\phi$ meson: QCD penguin topology in the full SM (left) and penguin contraction of the current$–-$current operator $\mathcal{O}_2^q$ in the low-energy effective Hamiltonian (right).}
    \label{fig:BsphiKSFeyn}
\end{figure}

However, in this case, the final states can factorize in two possible ways: the spectator quark can hadronize into either a $\bar{K}_{0}$ or a $\phi$ meson (see Fig.~\ref{fig:BsphiKSFeyn}.). The factorization can be written as
\begin{align}
\begin{split}          P^{(ct)}= \,\,&  \frac{G_F}{\sqrt{2}} \left[\tilde{a}_{\rm NF}^{K}\, \mathcal{C}_K\bra{\bar{K}^0}(\bar{d} s)_{\rm V-A}\ket{0}\bra{\phi}(\bar{s} b)_{\rm V-A}\ket{B_s^0} \right.\\ 
           +          & \left. \tilde{a}_{\rm NF}^{\phi}\, \mathcal{C}_\phi\bra{\phi}(\bar{s}s)_{\rm V-A}\ket{0}\bra{\bar{K}^0}(\bar{d} b)_{\rm V-A}\ket{B_s^0} \right],
\end{split}
\end{align}
where in the first term the $K^0$ decay constant enters and the term proportional to $\mathcal{C}_\phi$ factorizes analogously to the $B_d^0$ case, where the $\phi$ decay constant enters. However, the QCD penguin topologies contribute only to the first term. The $\tilde{a}_{\rm NF}^{K}$ and $\tilde{a}_{\rm NF}^{\phi}$ account for possible non-factorizable effects in the $b\to d$ transition. The relevant matrix elements related to $\mathcal{C}_K$ are parameterized as \cite{Isgur:1990kf}
\begin{align}
\label{eq:HMEK0}
    \bra{\bar{K}^0(p_P)}\bar{d}\gamma^\mu\gamma^5s\ket{0}&=if_K p_P^\mu\ , \\
    \label{eq:HMEphiBs}
    \bra{\phi(p_V,\epsilon)}\bar{s}\gamma^\mu\gamma^5b\ket{B_s^0(p_B)}&=i\left[a_+(q^2)P^\mu(\epsilon^* \cdot p_B)+a_-(q^2)q^\mu(\epsilon^* \cdot p_B)+f(q^2)\epsilon^{*\mu}\right],
\end{align}
where now $P\equiv p_B+p_V$, $q\equiv p_B-p_V=p_{P}$ and $f_K=(155.7\pm0.3)\,$MeV. The penguin operators give
\begin{align}\label{eq:CSM1}
   \mathcal{C}_K &= C_4+\frac{C_3}{3}-\frac{C_{10}+\frac{1}{3}C_{9}}{2}+
C_{\text{pen},c}^{\rm CC} \ ,
   \end{align}
 with $C_{\text{pen},c}^{\rm CC}$ defined in Eq.~\eqref{eq:CCpen}. For $\mathcal{C}_\phi$, we obtain 
\begin{equation}
    \mathcal{C}_\phi= C_3+C_5 +\frac{C_4+C_6}{3}-\frac{C_7+C_9+\tfrac{1}{3}C_{10}}{2}  -\frac{C_8}{6}\ .
\end{equation}
We note that $\mathcal{C}_K + \mathcal{C}_\phi = C_{c,\rm SM}$, where $\mathcal{C}_{c,\text{SM}}$ is given in Eq.~\eqref{eq:CSM}. Using the Wilson coefficients given in Ref.~\cite{Buras:1998raa} as before, we obtain
\begin{equation}
    \mathcal{C}_K=   [0.038,0.042]\ , \, \quad\quad \mathcal{C}_\phi = 0.002\ . 
\end{equation}
Here the variation of $ \mathcal{C}_K$ again stems from the $k^2/m_b^2$ variation in its physical range. As expected, $\mathcal{C}_\phi$, which is mediated by colour-allowed EW penguins and the colour-singlet contributions, is significantly smaller than $\mathcal{C}_K$. Hence, we only consider $\mathcal{C}_K$ contributions in the following.

Using the form factors in Eqs.~\eqref{eq:HMEK0} and \eqref{eq:HMEphiBs}, we write
\begin{align}
\begin{split}
        |P^{(ct)}|\propto \,&|\mathcal{C}_K|f_K F_1^{B_s^0 \phi}(q^2)\ ,
\end{split}
\end{align}
where the decay constant $f_K=(155.7\pm0.3)\,$MeV~\cite{FlavourLatticeAveragingGroupFLAG:2021npn}, and we defined~\cite{Melikhov:2000yu}
\begin{align}
    F_1^{B_s^0 \phi}(q^2)\equiv \frac{1}{2M_\phi}\left[f(q^2)+a_+(q^2)(M_{B_s^0}^2-M_\phi^2)+a_-(q^2) q^2\right].
\end{align}
Using the parameterization for $F_1^{B_s^0 \phi}(q^2)$~\cite{Melikhov:2000yu} detailed in Appendix~\ref{app:ff}, gives
\begin{align}
\label{eq:FFBsphi}
    F_1^{B_s^0\phi}(M_K^2)=0.43\pm0.04\ .
\end{align}
For the CP-averaged SM branching ratio expression of the $B_s\to\phi K_{\rm S}$ decay mode, we find 
\begin{equation}
\label{eq:BrBsSMtheo}
    \mathcal{B}(B_s^0\to \phi K_{\rm S})|_{\rm SM} = \frac{\hat{C}_s}{2} \left[1 - 2 \tilde{b} \cos\tilde{\theta}\cos\gamma+ \tilde{b}^2\right]  ,
\end{equation}
with 
\begin{equation}
    \hat{C}_s \equiv \tau_{B_s^0} \frac{G_F^2}{32\pi} A^2 \lambda^6 m_{B_s^0}^3 f_K^2 \left[\Phi\left(m_{K^0}/m_{B_s^0}, m_\phi/m_{B_s^0}\right)\right]^3\left[F^{B_s^0\phi}_1(m_{K^0}^2)\right]^2  |\mathcal{C}_{K}|^2 |a_{\rm NF}^s|^2 \ .
\end{equation}

The branching ratio above is a ``theoretical'' branching ratio. 
In $B_s^0$ systems, due to the sizeable decay width difference $\Delta \Gamma_s$, there is an important distinction between the ``theoretical'' branching ratio as defined above, and the ``experimental'' branching ratio. These quantities are related as follows \cite{DeBruyn:2012wj}:
\begin{equation}\label{eq:theorybr}
    \mathcal{B}(B_s^0 \to f)_{\rm theo} = \left[ \frac{1-y_s^2}{1+\mathcal{A}_{\Delta\Gamma}(B_s^0\to f) y_s}\right] \mathcal{B}(B_s^0\to f)_{\rm exp}\ .
    \end{equation}
 As noted, due to $y_d=\mathcal{O}(10^{-3})$, these effects play a negligible role in the $B_d^0$ system.

In the following section, we present predictions for SM branching ratios. However, before, let us further discuss the connection between the $\bsphiks$ and $\bdphiks$ modes. As already discussed, $P^{(ut)'}$ is expected to be of similar size as $P^{(ct)'}$ such that $b\sim R_b$. In the $\bdphiks$ and $\bplusphikplus$ decays, the effects of this parameter are strongly suppressed by $\epsilon$. Yet in $\bsphiks$, these effects are not suppressed. 
Despite the similar quark transitions between the $\bdphiks$ and $B_s^0\to\phi K_{\rm S}$ modes, there are no $U$-spin symmetry relations due to the different hadronization effects detailed above.

Despite these differences, the magnitudes of the hadronic penguin parameters of the  $\bdphiks$ and $B_s\to\phi K_{\rm S}$ modes are still expected to be comparable: 
\begin{equation}\label{eq:beqbtil}
    b \sim \tilde{b} \; \quad\quad \theta \sim \tilde\theta \ ,
\end{equation}
where any significant deviations could indicate NP contributions. As stressed, future measurements of the CP asymmetries in $\bsphiks$ could be used to determine $\tilde{b}$ and $\tilde\theta$ directly from the experimental data without any theoretical assumptions. These measurements would be highly welcome for further studies of the $B\to \phi K$ system.

\section{\texorpdfstring{ \boldmath SM predictions for $B_{(s)}\to \phi K$ decays}{}}
\label{sec:numerical}
\begin{figure}[t!]
    \centering
    \includegraphics[width=0.49\textwidth]{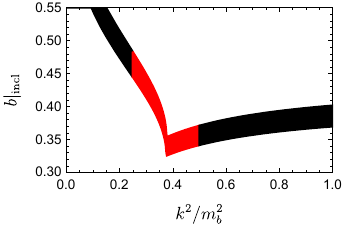}\includegraphics[width=0.49\textwidth]{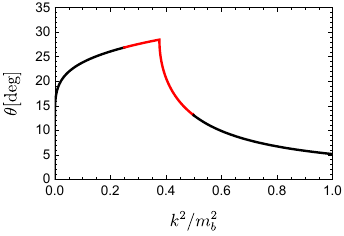}
    \caption{Hadronic parameters $b|_{\rm incl}$ and $\theta$ as functions of the squared momentum of the virtual gluon or photon. The red-shaded regions cover the allowed parameter ranges associated with the physical momentum range $1/4<k^2/m_b^2<1/2$~\cite{Gerard:1988jj}.}
    \label{fig:bthetadeltaP}
\end{figure}

We proceed by deriving SM estimates for the CP asymmetries and branching ratios of the $B\to \phi K$ decays utilizing the formalism given above. To this end, the penguin parameters $b$ and $\theta$ are an important input, which here we estimate using factorization arguments and kinematic considerations based on perturbative calculations of $\bar{c}c$ and $\bar{u}u$ loops, following the same approach as in Sec.~\ref{sec:fact}. In this approach, the only difference between $P^{(ct)'}$ and $P^{(ut)'}$ comes from the mass of the respective $c$ and $u$ light quarks, which run in the penguin loop. These enter in the penguin contraction of the current--current operators given by $\mathcal{C}_{\rm pen,q}$ defined in Eq.~\eqref{eq:CCpen}. Using factorization arguments, we already obtained the matrix element $P^{(ct)'}$ as given in Eq.~\eqref{eq:pct}. A similar expression for $P^{(ut)'}$ can be obtained by replacing $\mathcal{C}_{c, \rm SM}\to \mathcal{C}_{u,\rm SM}$. 
Using the expressions for $P^{(ct)'}$ and $P^{(ut)'}$, we then find
\begin{equation}
    b e^{i\theta} \equiv R_b \frac{P^{(ut)'}}{P^{(ct)'}} = R_b \frac{\mathcal{C}_{u, \rm SM}}{\mathcal{C}_{c,\rm SM}} \ .
\end{equation}
In the ``physical'' momentum range $1/4<k^2/m_b^2<1/2$, using again the Wilson coefficients from Ref.~\cite{Buras:1998raa}, we find 
    \begin{equation}
\label{eq:SMValuesWilsonVecu2}
    |\mathcal{C}_{u,\text{SM}}|  = [0.036, 0.040] \ .%0.039\pm0.003 \ ,
\end{equation}
We note that, as expected, this result is similar to the value of $\mathcal{C}_{c,\rm SM}$ given in Eq.~\eqref{eq:SMValuesWilsonVec2}. Using the inclusive $R_b$ as an example, we show in Fig.~\ref{fig:bthetadeltaP}, the dependence of $b|_{\rm incl}$ on the phase $\theta$ as functions of the normalized momentum $k^2/m_b^2$, also beyond the ``physical'' region. The band in Fig.~\ref{fig:bthetadeltaP} arises from the uncertainty on $R_b$. We observe that the physical range covers a large part of all allowed values. From the distribution of the penguin parameters over the physical momentum range $1/4<k^2/m_b^2<1/2$, we obtain 
\begin{align}
\label{eq:hadrparRanges22}
    b|_{\rm incl}=0.38\pm0.04,\quad b|_{\rm excl}=0.34\pm 0.04,\quad \theta=(23\pm 6)^\circ\,,
\end{align} 
for the inclusive and exclusive values for $R_b$, respectively and, $\theta$ is essentially independent of $R_b$.  We show the correlation between $b|_{\rm incl}$ and $\theta$ in Fig.~\ref{fig:DeltaPeffphidAdir}. Here, the theoretical uncertainty bands are obtained by varying the virtual gluon and photon momenta within the $k^2/m_b^2\in[0.1,1]$ range, and values within the physical range are highlighted in red. We find $b\sim R_b$, indicating that the perturbative contribution to $P^{ut'}$ and $P^{ct'}$ are roughly equal. 
At the same time, we note that the above results are estimates.
In addition, the expression for the loop function in Eq.~\eqref{eq:Gmu} is a simplification instead of an integration over the wave functions of the final state mesons. By using the physical $k^2$ range, we account for the uncertainty this assumption may introduce.

The mixing-induced CP asymmetry in the $\bdphiks$ mode is proportional to $\sin\phi_d$ with small penguin effects. Therefore, it is interesting to compare this observable with its counterpart in $B_d^0\to J/\psi K_S$ decays, which exhibits the same dependence. We define 
\begin{align}\label{eq:Deltapsiphidef}
    \Delta\mathcal{A}^{\rm mix}_{\rm \psi \phi}& \equiv \mathcal{A}^{\rm mix}_{\rm CP}(B_d^0\to J/\psi K_S) -     \mathcal{A}^{\rm mix}_{\rm CP}(B_d^0\to \phi K_S)   \ ,
\end{align}
where the leading terms proportional to $\sin\phi_d$ cancel. We stress that, therefore, also possible CP-violating NP effects in $B_d^0$--$\bar{B}_d^0$ mixing cancel in the difference. In the SM, we find
\begin{align}\label{eq:Deltapsiphi}
    \Delta\mathcal{A}^{\rm mix}_{\rm \psi \phi}|_{\rm SM}& 
    = 2\epsilon \sin \gamma \cos \phi_d (b \cos \theta - a^\prime \cos \theta^\prime ) + \mathcal{O}(\epsilon^2)\ ,
\end{align}
where $a^\prime$ and $\theta^\prime$ are the penguin parameters of the $B_d^0\to J/\psi K_S$ decay \cite{DeBruyn:2014oga}, analogous to $b$ and $\theta$. In the $B_d^0\to J/\psi K_S$ decay, the penguin parameters can be extracted from data using $U$-spin and $SU(3)$ control channels. Taking the values obtained in Ref.~\cite{DeBruyn:2014oga}, we find
\begin{align}
    \Delta\mathcal{A}^{\rm mix}_{\rm \psi \phi}|_{\rm incl}=0.037\pm 0.010, \quad \Delta\mathcal{A}^{\rm mix}_{\rm \psi \phi}|_{\rm excl}=0.035\pm 0.010.
\end{align}
Using $\mathcal{A}^{\rm mix}_{\rm CP}(B_d^0\to J/\psi K_S)|_{\rm exp}=-0.710\pm0.014$ \cite{ParticleDataGroup:2024cfk}, the current experimental data give 
\begin{align}\label{eq:psiphiexp}
    \Delta\mathcal{A}^{\rm mix}_{\rm \psi \phi}|_{\rm exp}=-0.13\pm0.12, 
\end{align}
which leaves room for NP effects at decay amplitude level of $~20\%$. We further discuss these possible NP effects in Sec.~\ref{sec:NewPhysics}. 

\subsection{\texorpdfstring{ \boldmath $B_d^0\to\phi K_{\rm S}$ and $B^+\to\phi K^+$ in the SM}{}}

Using the hadronic penguin parameters $b$ and $\theta$ in Eq.~\eqref{eq:hadrparRanges22}, we may calculate the CP asymmetries and hadronic phase shift $\Delta\phi_d$ for the $\bdphiks$ decay within the SM. For the observables of the $\bdphiks$ decay introduced in Eqs.~\eqref{eq:AdirBdksphi}--\eqref{eq:ADeltaGammaBdksphi}, we find 
\begin{align}
\label{eq:AdirTheo}
        \mathcal{A}^{\rm dir}_{\rm CP}(\bdphiks)|_{\rm incl}&= -0.014\pm0.005,\quad &&\,\mathcal{A}^{\rm dir}_{\rm CP}(\bdphiks)|_{\rm excl}=-0.012\pm0.004,\\
        \mathcal{A}^{\rm mix}_{\rm CP}(\bdphiks)|_{\rm incl}&=-0.723\pm0.019,\quad &&\mathcal{A}^{\rm mix}_{\rm CP}(\bdphiks)|_{\rm excl}=-0.721\pm0.019,\\
        \mathcal{A}^{\Delta\Gamma}_{\rm CP}(\bdphiks)|_{\rm incl}&= \phantom{-}0.690\pm0.020,\quad &&\mathcal{A}^{\Delta\Gamma}_{\rm CP}(\bdphiks)|_{\rm excl}= \phantom{-}0.693\pm0.020.
\end{align}

We remind the reader that these observables are not independent from one another and that $\mathcal{A}^{\Delta\Gamma}_{\rm CP}$ will be difficult to measure experimentally due to the small width difference $\Delta \Gamma_d$. As discussed, our predictions of the direct CP asymmetry for the charged $\bplusphikplus$ mode are equal to that of $\bdphiks$ in the SM within our approximations. Our estimates for the direct CP-asymmetry fall within the experimental values listed in Tab~\ref{tab:expvalues} at the $1\sigma$ level. On the other hand, for the mixing-induced CP asymmetry, we find predictions that are $1\sigma$ above the current experimental result. Therefore, updated measurements, especially of mixing-induced CP asymmetries, would be very interesting for further understanding of the penguin parameters. 

For the theoretical values of the hadronic phase shift $\Delta\phi_d^{\phi K_{\rm S}}$ in Eq.~\eqref{eq:hadronTan}, we estimate 
\begin{align}
\label{eq:deltaphism}
\Delta\phi_d^{\phi K_{\rm S}}|_{\text{incl}}= (1.9\pm0.2)^\circ \ ,  \quad \Delta\phi_d^{\phi K_{\rm S}}|_{\text{excl}}= (1.7\pm0.2)^\circ \ ,
\end{align}
which have a $10\%$ relative uncertainty. As discussed, the current experimental data on the mixing-induced CP asymmetry still have large uncertainties. From Eq.~\eqref{eq:Defphidef}, we find
\begin{equation}
    \phi_{d,\phi K_{\rm S}}^{\rm eff} = (34 \pm 8 )^\circ \ ,
\end{equation}
which is more than $1\sigma$ below the experimental value for $\phi_d$. At the same time, comparing with the hadronic penguin effects above, which are about $\mathcal{O}(2\degree)$, we note that the precision on $\phi_{d, \phi K_{\rm S}}^{\rm eff}$ is about a factor of four larger. Consequently, once future measurements approach a precision below the $5\degree$ level on $\phi_{d, \phi K_{\rm S}}^{\rm eff}$, it will be important to include these hadronic phase shifts. Using our input for $\phi_d$, we can convert $\phi_{d, \phi K_{\rm S}}^{\rm eff}$ into a constraint on CP-violating NP at the decay amplitude level. We find 
\begin{equation}
    \Delta \phi_d^{\rm NP} = (-12\pm 8)^\circ \ ,
\end{equation}
which shows significant room for such NP. In the Sec.~\ref{sec:NewPhysics}, we discuss such effects in more detail.

\begin{figure}
    \centering
    \includegraphics[width=0.495\textwidth]{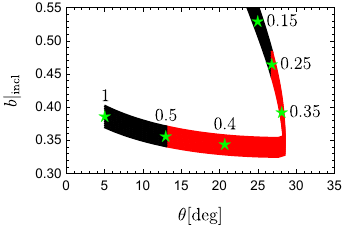}\includegraphics[width=0.495\textwidth]{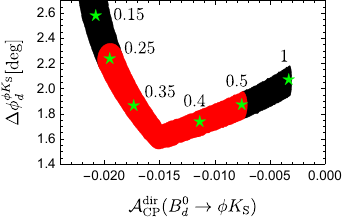}
    \caption{Correlations between the $\bdphiks$ hadronic penguin parameters (left), and the direct CP asymmetry with the hadronic phase shift (right), obtained by varying the virtual gluon and photon momenta over the range $0<k^2<m_b^2$. The red regions indicate the physical range $1/4<k^2/m_b^2<1/2$, while the green stars denote momentum fractions $k^2/m_b^2 = \{1,0.5, 0.4, 0.35, 0.25, 0.15\}$, as indicated.}
    \label{fig:DeltaPeffphidAdir}
\end{figure}

Anticipating these new measurements, we show the correlation between the direct CP asymmetry and the hadronic phase shift in the SM in Fig.~\ref{fig:DeltaPeffphidAdir} for the $k^2/m_b^2 = [0.1,1]$ range. The physical region for the momentum fraction is again highlighted in red. Should future experimental results lie outside the shaded regions, this would be very interesting, as it could indicate NP contributions. 

While the CP asymmetries only depend on the hadronic parameters $b, \theta$, the $\bdphiks$ branching ratio in Eq.~\eqref{eq:Pctex} also involves the overall penguin amplitude $P^{(ct)\prime}$. We obtain the penguin amplitude using the factorization estimate in Eq.~\eqref{eq:BRstandardModel} and provide predictions for both the inclusive and exclusive determinations of the relevant CKM matrix elements. For completeness, we also present them in the limit where the hadronic penguin parameters are set to zero. This requires the $B\to K^0$ form factor as input. We use the Lattice QCD determination from \cite{Bourrely:2008za,FlavourLatticeAveragingGroupFLAG:2021npn} (see  Appendix~\ref{app:ff}), which reads  
\begin{equation}\label{eq:FB0K0}
    F^{B^0K^0}_1(m_\phi^2)=0.391\pm0.030\ ,
\end{equation} 
and has a sizeable $\mathcal{O}(10\%)$ uncertainty. 

For the estimates of the hadronic parameters $b$ and $\theta$ in Eq.~\eqref{eq:hadrparRanges22}, we then find 
\begin{align}
\label{eq:BtheoOverC2a2}
    \frac{\mathcal{B}(B_d^0 \to \phi K_{\rm S})|_{\rm theo}^{\rm SM}}{|\mathcal{C}_{c,\rm SM}|^2(a_{\rm NF}^0)^2}\times 10^4=\begin{cases}
    \text{Incl}: \;28.1\pm4.7, \quad \;{\rm Incl}|_{b = 0}: \;27.7\pm4.6 \\
    \text{Excl}: 25.0\pm4.1, \quad {\rm Excl}|_{b = 0}: \;24.6\pm4.1\ ,
    \end{cases}
\end{align}
and
\begin{align}
\label{eq:BptheoOverC2a2}
    \frac{\mathcal{B}(\bplusphikplus)|_{\rm theo}^{\rm SM}}{|\mathcal{C}_{c,\rm SM}|^2(a_{\rm NF}^+)^2}\times 10^4=\begin{cases}
    \text{Incl}: \;60.7\pm10.0, \quad \;{\rm Incl}|_{b = 0}: \;59.7\pm9.9 \\
    \text{Excl}: 53.9\pm8.9, \quad \;\;{\rm Excl}|_{b = 0}: 53.2\pm8.8\ .
    \end{cases}
\end{align}

Assuming ideal factorization, i.e., setting $\delta_{\rm NF}^{0,+}=0$ such that $a_{\rm NF}^{0,+}=1$, and using Eq.~\eqref{eq:SMValuesWilsonVec2}, we find
\begin{align}
\label{eq:brsmpred}
\mathcal{B}(\bdphiks)|^{\rm fact}_{\rm SM}\times  10^{6}=
\begin{cases}
    \text{Incl}: \;4.9 \pm 1.1, \quad \;{\rm Incl}|_{b = 0}: \;4.8 \pm 1.1 \\
    \text{Excl}: 4.3 \pm 1.0, \quad {\rm Excl}|_{b = 0}: \;4.3 \pm 1.0\ ,
    \end{cases}
\end{align}   
and
\begin{align}
\label{eq:brsmpredbp}
\mathcal{B}(\bplusphikplus)|^{\rm fact}_{\rm SM}\times  10^{6}=
\begin{cases}
    \text{Incl}: \;10.5\pm2.4, \quad \;{\rm Incl}|_{b = 0}: \; 10.4\pm2.4\\
    \text{Excl}: 9.4\pm2.2 , \quad {\rm Excl}|_{b = 0}: \; 9.2\pm2.1\ .
    \end{cases}
\end{align}   
The relatively large uncertainty arises from the $10\%$ uncertainty on the form factor and a similar uncertainty on $\mathcal{C}_{c,\rm SM}$. We note that the doubly Cabibbo-suppressed penguin parameters $b$ and $\theta$ have a negligible effect on the branching ratio. In addition, we note a $\mathcal{O}(10\%)$ discrepancy between branching ratio predictions for the inclusive and exclusive CKM factors. 

To compare these estimates with the measurements, we define
\begin{align}
\label{eq:DelB}
   \Delta \mathcal{B}(B\to\phi K) \equiv  \frac{ \mathcal{B}(B\to\phi K)|_{\rm exp.}}{\mathcal{B}(B\to\phi K)|^{\rm fact}_{\rm SM}}\ 
\end{align}
with $B=B_d^0,B^+,K=K_{\rm S},K^+$. Assuming factorization, we find
\begin{align}
    \Delta \mathcal{B}(B_d^0\to\phi K_{\rm S})=\begin{cases}
    \text{Incl}: \;0.75 \pm 0.19, \quad \;{\rm Incl}|_{b = 0}: \;0.76 \pm 0.19 \\
    \label{eq:DelBKSexcl}
    \text{Excl}: 0.84 \pm 0.21, \quad {\rm Excl}|_{b = 0}: \;0.85 \pm 0.21 \ ,
    \end{cases}
\end{align}
and
\begin{align}\label{eq:DelBplus}
    \Delta \mathcal{B}(\bplusphikplus)=\begin{cases}
    \text{Incl}: \; 0.84\pm0.20, \quad \;{\rm Incl}|_{b = 0}: \; 0.95\pm0.23 \\
    \text{Excl}: 0.94\pm0.23, \quad {\rm Excl}|_{b = 0}: \; 0.85\pm0.21 \ .
    \end{cases}
\end{align}
We find a pattern where the measured branching ratios are below our predictions by $1$--$2\sigma$. Given the simplicity of the factorization framework, this agreement with the data is encouraging. 

It is therefore interesting to turn the argument around and directly extract the size of the non-factorizable effects from the data within the SM. For the SM predictions of $\delta_{\rm NF}^{0,+}$, using the experimental branching ratios in Tab.~\ref{tab:expvalues}, we find
\begin{align}
\label{eq:anfes}
    \delta_{\rm NF}^0=\begin{cases}
    \text{Incl}: \;-0.14\pm 0.12, \quad \;{\rm Incl}|_{b = 0}: \;-0.13\pm 0.12 \\
    \text{Excl}: -0.08\pm 0.12, \quad {\rm Excl}|_{b = 0}: \;-0.08\pm 0.12 \ ,
    \end{cases} 
\end{align}
and
\begin{align}
\label{eq:anfesbp}
    \delta_{\rm NF}^+=\begin{cases}
    \text{Incl}: \; -0.09\pm0.12, \quad \;{\rm Incl}|_{b = 0}: \; -0.08\pm0.12 \\
    \text{Excl}: -0.03\pm0.12, \quad {\rm Excl}|_{b = 0}: \; -0.02\pm0.12 \ .
    \end{cases}
\end{align}
This shows that describing the perturbative effects using the simple estimate in Eq.~\eqref{eq:CSM} yields remarkably small non-factorizable contributions, supporting the applied framework. 
At the same time, there remains ample room for both improvements on the theoretical side and potential NP contributions.

\subsection{\texorpdfstring{ \boldmath $B_s^0\to\phi K_{\rm S}$ in the SM}{}}
In the following, we assume that the penguin parameters in $B_s^0\to \phi K_{\rm S}$ are similar to those in $B_d\to \phi K_{\rm S}$. Using the estimates for the penguin parameters in Eq.~\eqref{eq:hadrparRanges22}, we obtain
\begin{align}
    \mathcal{A}_{\text{CP}}^{\text{dir}}(B_s^0\to \phi K_{\rm S})|_{\rm incl}&=0.32\pm0.10,\quad  &&\,\mathcal{A}_{\text{CP}}^{\text{dir}}(B_s^0\to \phi K_{\rm S})|_{\rm excl}=0.28\pm0.09,\\ \mathcal{A}_{\text{CP}}^{\text{mix}}(B_s^0\to \phi K_{\rm S})|_{\rm incl}&=0.64\pm0.04,\quad &&\mathcal{A}_{\text{CP}}^{\text{mix}}(B_s^0\to \phi K_{\rm S})|_{\rm excl}=0.59\pm0.04,\\
    \mathcal{A}^{\Delta\Gamma}_{\rm CP}(B_s^0\to \phi K_{\rm S})|_{\rm incl}&=0.69\pm0.07,\quad &&\,\mathcal{A}^{\Delta\Gamma}_{\rm CP}(B_s^0\to \phi K_{\rm S})|_{\rm excl}=0.75\pm0.05.
\end{align}
% tabulated in Tab.~\ref{tab:theoretical}.

Measuring the direct and mixing-induced CP asymmetries in this decay mode is challenging, as it requires information on the timing and tagging of the $B_s$ mesons. On the other hand, the untagged observable $\mathcal{A}_{\rm CP}^{\Delta \Gamma}(B_s^0\to\phi K_{\rm S})$ only requires time information and is therefore more easily accessible. Future analyses of this observable and CP violation in $\bsphiks$ can test our SM predictions and give insights into the size of the hadronic contributions.

\begin{figure}[t!]
    \centering  \includegraphics[width=0.495\textwidth]{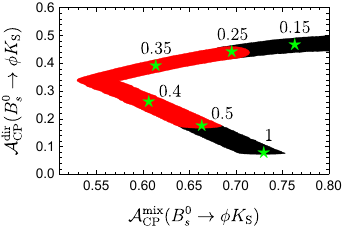}\includegraphics[width=0.495\textwidth]{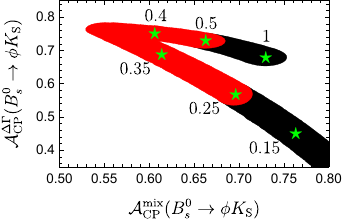}
    \caption{Correlations of the mixing-induced CP asymmetry of the $\bsphiks$ decay with the direct CP asymmetry (left) and with $\mathcal{A}_{\rm CP}^{\Delta\Gamma}$ (right). The scan over the virtual gluon and photon momenta and the graphical conventions are the same as in Fig.~\ref{fig:DeltaPeffphidAdir}.}
    \label{fig:DeltaGammacorrBsphiks}
\end{figure}

In Fig.~\ref{fig:DeltaGammacorrBsphiks}, we present the correlation between $\mathcal{A}^{\Delta\Gamma}_{\rm CP}$ and the direct and mixing-induced CP asymmetries in $\bsphiks$ decays, using the same procedure and graphical conventions as in Fig.~\ref{fig:DeltaPeffphidAdir}. As before, future measurements outside the shaded region could indicate NP contributions. 

For the theoretical branching ratio estimates, we assume $a_{\rm NF}^s = a_{\rm NF}^0 = 1+\delta^0_{\rm NF}$ for the non-factorizable effects. This leads to 
\begin{align}
\mathcal{B}(B_s^0\to \phi K_{\rm S})|_{\rm SM,theo}\times 10^8=
\begin{cases}
&\tilde{b} = b:\;8.6\pm3.1\\ &\tilde{b} = 0: \; 10.3\pm 3.6 \ .
    \end{cases}
\end{align}

Using Eq.~\eqref{eq:theorybr}, the estimates for the ``experimental'' $\bsphiks$ branching ratio are 
\begin{align}
\mathcal{B}(B_s^0\to \phi K_{\rm S})|_{\rm SM,exp}\times 10^8=
\begin{cases}
&\tilde{b} = b:\;\;\;9.0\pm3.3\\ &\tilde{b} = 0: \; 10.8\pm 3.8 \ .
    \end{cases} 
\end{align}
As we pointed out, the $B_s\to \phi K_{\rm S}$ channel is of particular interest as the penguin parameters $\tilde{b}$ and $\tilde\theta$ do not enter with the doubly Cabibbo-suppressed parameter $\epsilon$, in contrast to the situation in the $\bdphiks$ and $\bplusphikplus$ decays. The two determinations above show the strong sensitivity of the branching ratio to the hadronic penguin contributions, highlighting that already the first determinations of the branching ratio would be very valuable for our understanding of the $B\to \phi K$ system.

\section{Isospin-dependent observables}
\label{sec:isospinSM}
As already noted, the dominant modes of the neutral and charged modes $B\to\phi K$ are related through isospin symmetry. In the isospin limit, neglecting suppressed annihilation and exchange contributions, the direct CP asymmetries of the neutral and charged modes are equal as noted in Eq.~\eqref{eq:adirequal}. The $B_d^0\to\phi K^0$ and $\bplusphikplus$ decay amplitudes are similarly identical in the SM. The isospin-breaking effects in the phase-space factors of the branching ratios in these modes are \cite{ParticleDataGroup:2024cfk}
\begin{align}\label{eq:Xisotheo}
X_{\rm isospin} \equiv
\frac{\tau_{B_d^0}}{\tau_{B^+}}
\frac{M_{B^+}}{M_{B_d^0}}
\frac{\Phi\left( m_{K^0}/m_{B_d^0}, m_\phi/m_{B_d^0}\right)}
{\Phi\left(m_{K^+}/{m_{B^+}}, m_\phi/m_{B^+}\right)}
=0.926\pm0.03,
\end{align}
which is a $7\%$ effect. Neglecting isospin-breaking spectator-quark effects, we then expect
\begin{equation}
\mathcal{B}(B_d^0\to \phi K^0)
= 2\, \mathcal{B}(\bdphiks)
= X_{\rm isospin}\, \mathcal{B}(\bplusphikplus)\ .
\end{equation} Using the experimental branching ratios listed in Table~\ref{tab:expvalues}, we obtain
\begin{align}\label{eq:Xisospinexp}
\Delta\mathcal{B}_{\rm isospin}
=\frac{\mathcal{B}(B_d^0\to\phi K^0)}
{\mathcal{B}(\bplusphikplus)}
=0.83\pm 0.10,
\end{align}
which agrees within $1\sigma$ with the isospin limit. 

We thus find that the current $B\to\phi K$ data are consistent with our SM estimates. However, given the present experimental uncertainties, substantial room remains for isospin-breaking SM effects and for possible contributions from NP. Identifying such effects is a key motivation for studying decay channels governed by penguin topologies. In the following, we introduce isospin-dependent observables and perform a SM analysis.

Following Refs.~\cite{Fleischer:2001pc, Fleischer:2001cw}, we decompose the low-energy effective Hamiltonian in Eq.~\eqref{eq:Heff} as 
\begin{equation}
    \mathcal{H}_{\rm eff} = \mathcal{H}_{\rm eff}^{I=0}+\mathcal{H}_{\rm eff}^{I=1} \ ,
\end{equation}
taking into account that the combinations $(u\bar{u}+ d\bar{d})$ and $(u\bar{u}-d\bar{d})$ are associated with isospin $I=0$ and $I=1$ transitions, respectively. Accordingly, the penguin amplitudes $P^{(qt)'}$ in Eqs.~\eqref{eq:apprime} and~\eqref{eq:hadronicParameters} can be decomposed into their $I=0$ and $I=1$ contributions as 
\begin{align}
\label{eq:decayampiso}
    A(B^{0,+}\to\phi K^{0,+})&=\frac{\mathcal{A}_{P,\pm}^\prime}{\sqrt{\epsilon}}\left[1+\epsilon b_\pm e^{i\theta_\pm}e^{i\gamma}\right], 
\end{align}
where the subscripts on the right-hand side signify the $\bar{u}u\pm\bar{d}d$ flavour structure and the upper (lower) sign holds for $B_d^0 (B^+)$. We defined 
\begin{equation}
    \mathcal{A}_{P,\pm}^\prime \equiv A\lambda^3 (P^{(ct)\prime}_{(0)}\pm P^{(ct)\prime}_{(1)}) \ ,
\end{equation}
as well as
\begin{equation}
    b_{\pm } e^{\pm i\theta_\pm}\equiv R_b \left[\frac{P^{(ut)'}_{(0)} \pm P^{(ut)'}_{(1)}}{P^{(ct)'}_{(0)} \pm P^{(ct)'}_{(1)}}\right] \ .
\end{equation}
We stress that the $P^{(ct)'}_{(1)}$ amplitude is essentially generated only through EW penguins, which are suppressed by $\alpha/\alpha_s$. 

Following Ref.~\cite{Fleischer:2001pc}, we also take into account the dynamical suppression of the corresponding hadronic matrix element and estimate
\begin{equation}\label{eq:1over0}
    |P^{(ct)'}_{(1)}/P^{(ct)'}_{(0)}| \sim \mathcal{O}(\overline{\lambda}^2) \ , 
\end{equation}
where $\overline{\lambda}=0.2$ \cite{Gronau:1995hn,Buras:1995pz} is a generic expansion parameter of the same order as the Wolfenstein parameter $\lambda$.
Therefore, we find 
\begin{equation}\label{eq:applusmin}
     \mathcal{A}_{P,+}^\prime =  \mathcal{A}_{P,-}^\prime 
    \left[1 + \mathcal{O}(\overline{\lambda}^2)\right]\ .
\end{equation}
Assuming that the penguin amplitudes are dominated by top quarks in the loop, we also expect $P^{(ut)^\prime}$ to scale as in Eq.~\eqref{eq:1over0}. We thus approximate the difference between the hadronic parameters as 
\begin{align}\label{eq:best}
    b_+-b_-=R_b\left[\frac{1+\mathcal{O}(\overline{\lambda}^2)}{1+\mathcal{O}(\overline{\lambda}^2)}-\frac{1-\mathcal{O}(\overline{\lambda}^2)}{1-\mathcal{O}(\overline{\lambda}^2)}\right]=\mathcal{O}(\overline{\lambda}^2)\ .
\end{align}

Using these estimates, we can further quantify the differences between the $B^0_d\to\phi K^0$ and $\bplusphikplus$ branching ratios. To distinguish between the isospin states, we write the SM branching ratio expressions as 
\begin{align}\label{eq:BSM}
    \mathcal{B}\left(B^{0,+}\to \phi K^{0,+}\right) &= \frac{\hat{C}^{0,+}}{\epsilon} \left[1+2\epsilon b_\pm \cos\theta_\pm \cos\gamma+\left(\epsilon b_\pm\right)^2\right]\ ,
\end{align}
where we have already defined $\hat{\mathcal{C}}^{0,+}$ around Eq.~\eqref{eq:Chatdef}. The subscripts of the penguin parameters $b$ and $\theta$ again signify the sign of the $\bar{u}u\pm\bar{d}d$ flavour structures. We now write
\begin{equation}\label{eq:UtimesoneplusW}
    \mathcal{U}(1+\mathcal{W}) = \frac{  \mathcal{B}\left(B^{+}\to \phi K^{+}\right)} {\mathcal{B}\left(B^{0}_d\to \phi K^{0}\right)} \frac{\hat{\mathcal{C}}^0}{\hat{\mathcal{C}}^+} \ ,
\end{equation}
where we have defined
\begin{align}
\label{eq:mathcalU}
\mathcal{U}\equiv\displaystyle\frac{\mathcal{B}(B^+\to\phi K^+)}{\mathcal{B}(B_d^0\to\phi K^0)}\frac{\tau_{B_d^0}}{\tau_{B^+}} \frac{m_{B_d^0}^3}{m_{B^+}^3}\left[\frac{\Phi\left(\frac{m_{K^0}}{m_{B_d^0}},\frac{m_\phi}{m_{B_d^0}}\right)}{\Phi\left(\frac{m_{K^+}}{m_{B^+}},\frac{m_\phi}{m_{B^+}}\right)}\right]^3\ ,
\end{align}
and
\begin{equation}\label{eq:oneplusW}
    (1+\mathcal{W})\equiv \left[\frac{F^{B^0K^0}_1(m_\phi^2)}{F^{B^+K^+}_1(m_\phi^2)}\right]^2  \left[\frac{a_{\rm NF}^0}{a_{\rm NF}^+}\right]^2 \ .
\end{equation}
The quantity $\mathcal{U}$ captures the isospin corrections to the phase-space factors, whereas $\mathcal{W}$ quantifies total non-factorizable isospin-breaking effects as well as the isospin corrections to the form factors in the factorization approach. In the SM, we find in terms of hadronic penguin parameters
\begin{align}\label{eq:UtimesoneplusWSM}
    \mathcal{U}(1+\mathcal{W})|_{\rm SM}  &= \frac{1+2\epsilon b_+ \cos\theta_+\cos\gamma+\left(\epsilon b_+\right)^2}{1+2\epsilon b_- \cos\theta_-\cos\gamma+\left(\epsilon b_-\right)^2} \\ &= 1 + 2 \epsilon \cos\theta \cos\gamma (b_+-b_-) + \mathcal{O}(\epsilon^2)\label{eq:UtimesoneplusWSM2}  \ , 
\end{align}
where we have assumed for the CP-conserving strong phases $\theta_+= \theta_-\equiv \theta$ in Eq.~\eqref{eq:UtimesoneplusWSM2}. 

Since isospin-breaking corrections are typically expected to be at the percent level, we express Eq.~\eqref{eq:oneplusW} in terms of small isospin-breaking parameters $\epsilon_F$ and $\epsilon_a$ defined through
\begin{align}
\frac{F^{B^0K^0}}{F^{B^+ K^+}_1}&=
1+\epsilon_F+\mathcal{O}(\epsilon_F^2)\ ,\\
\frac{a_{\rm NF}^0}{a_{\rm NF}^+} &= 1+\epsilon_a+\mathcal{O}(\epsilon_a^2)\ .
\end{align}
Finally, neglecting terms quadratic in $\epsilon_F$ and $\epsilon_a$ leads to 
 \begin{equation}
 \label{eq:W4pct}
     \mathcal{W} = 2\epsilon_F + 2 \epsilon_a\ .
 \end{equation}
We treat these two corrections separately because $\epsilon_F$ could -- in principle -- be obtained from Lattice or Light-Cone Sum Rule (LCSR) QCD methods. We note that $\epsilon_a$ also contains weak annihilation contributions present in the $B^+$ decay, but not in the $B_d^0$ case. Allowing for a typical isospin correction of at the $1\%$ level in both terms implies $\mathcal{W}=\mathcal{O}(4\%)$. 

In order to probe the level of isospin-breaking effects in the hadronic penguin parameters,  we can use the differences in the normalized branching ratios and define
\begin{align}
    \label{eq:ZobsUW}
    \mathcal{Z}&\equiv\frac{1-\mathcal{U}\left(1+\mathcal{W}\right)}{1+ \mathcal{U}\left(1+\mathcal{W}\right)} .
\end{align}
Assuming the SM, using Eq.~\eqref{eq:best} and neglecting $\epsilon^2$ terms, this expression reduces to 
\begin{equation}\label{eq:ZSM}
\mathcal{Z}_{\rm SM} = \epsilon \cos \theta \cos \gamma  (b_- - b_+) = \mathcal{O}(\lambda^2 \times\overline{\lambda}^2) \simeq \mathcal{O}(0.1\%)\ ,
\end{equation}
which is remarkably suppressed. In principle, rescattering effects may enhance the estimate in Eqs.~\eqref{eq:1over0} and \eqref{eq:best} by $\mathcal{O}(\overline{\lambda})$, as long-distance final-state interactions can relax the dynamical suppression of the $I=1$ contributions assumed in the short-distance framework. Therefore, to be conservative, we increase this estimate by an order of magnitude and use $\mathcal{Z}_{\rm SM}=\mathcal{O}(1\%)$.

Finally, we can construct two more isospin-dependent observables employing the direct CP asymmetries through
\begin{align}\label{eq:SandD}
       \mathcal{S}&\equiv \frac{1}{2} \left[\mathcal{A}_\text{CP}^{\text{dir}}(B_d^0\to\phi K^0)+\mathcal{A}_\text{CP}^{\text{dir}}(B^+\to\phi K^+)\right],\nonumber \\
       \mathcal{D}&\equiv \frac{1}{2} \left[\mathcal{A}_\text{CP}^{\text{dir}}(B_d^0\to\phi K^0)-\mathcal{A}_\text{CP}^{\text{dir}}(B^+\to\phi K^+)\right].
\end{align}
Interestingly, in the presence of isospin-dependent NP, $\mathcal{S}$ is sensitive to $I=0$ NP contributions, while $\mathcal{D}$ and $\mathcal{Z}$ probe the $I=1$ sector \cite{Fleischer:2001pc, Fleischer:2001cw}. We discuss this further in the following section.

From Eqs.~\eqref{eq:gamma} and \eqref{eq:UtimesoneplusWSM}, and taking $\theta_+=\theta_-=\theta=(23\pm6)^\circ$ as given in Eq.~\eqref{eq:hadrparRanges22}, we obtain
\begin{equation}\label{eq:UtimesoneplusWSM3}
    \mathcal{U}(1+\mathcal{W})|_{\rm SM}  \sim  1+ (0.041\pm 0.007) \times (b_+-b_-)\ . 
\end{equation}
Using the experimental inputs in Tab.~\ref{tab:expvalues} yields
\begin{align}
\label{eq:mathcalUval}
\mathcal{U}=1.07\pm0.13\ .
\end{align}
This result is consistent with the isospin-symmetry limit for the spectator quarks, $\mathcal{U}|_{\rm isospin}=1$. However, the $10\%$ experimental uncertainty on the branching ratios still allows for sizeable differences between $b_+$ and $b_-$. 

\begin{figure}[t!]
    \centering
    \includegraphics[width=0.55\textwidth]{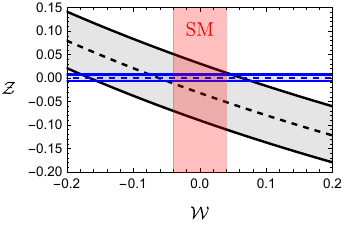}
    \caption{\small Current experimental bounds on $\mathcal{Z}$ (black), defined in Eq.~\eqref{eq:ZobsUW}, together with the SM prediction (blue) given in Eq.~\eqref{eq:ZSM}. The red band highlights the SM expectation $\mathcal{W}=\mathcal{O}(4\%)$.}
    \label{fig:MannelBX}
\end{figure}

In Fig.~\ref{fig:MannelBX}, we show $\mathcal{Z}$ as a function of $\mathcal{W}$, using the experimental input for $\mathcal{U}$ from Eq.~\eqref{eq:mathcalUval}. We compare this with the SM prediction for $\mathcal{Z}$ given in Eq.~\eqref{eq:ZSM}. The current experimental constraint on $\mathcal{U}$ implies that values in the range $-0.19 < \mathcal{W} < 0.05$ remain compatible with $\mathcal{Z}_{\rm SM}$. This interval is consistent with our SM estimate $\mathcal{W}=\mathcal{O}(4\%)$ obtained in Eq.~\eqref{eq:W4pct}. Still, it is quite remarkable that we can obtain information on these isospin-breaking observables from such experimentally challenging hadronic decays.   

On the other hand, adopting the SM estimate for $\mathcal{W}$ allows us to extract $\mathcal{Z}$ from the experimental data. This results in
\begin{align}
\label{eq:mathcalZlimit}
    \mathcal{Z}|_{\mathcal{W}=\mathcal{O}(4\%)}=-0.03\pm0.07\ ,
\end{align}
which is consistent with the SM prediction for $\mathcal{Z}$ provided in Eq.~\eqref{eq:ZSM}. 
We therefore conclude that current data exhibit no significant deviations from the SM estimates. 
It will nevertheless be important to monitor future improvements in the theoretical uncertainty of $\mathcal{W}$, as well as the experimental bounds on $\mathcal{U}$. If consistency between $\mathcal{Z}$ in Eq.~\eqref{eq:ZobsUW} and the SM prediction $\mathcal{Z}_{\rm SM}$ in Eq.~\eqref{eq:ZSM} can only be achieved for values of $\mathcal{W}$ significantly outside the SM estimate in Eq.~\eqref{eq:W4pct}, this would signal isospin-dependent NP contributions. We elaborate on this in Sec.~\ref{sec:NewPhysics}.

Imposing the current experimental constraints in Tab.~\ref{tab:expvalues}, together with $\mathcal{A}_\text{CP}^{\text{dir}}(B_d^0\to\phi K^0)=\mathcal{A}_\text{CP}^{\text{dir}}(B_d^0\to\phi K_{\rm S})$, we find
\begin{align}
\label{eq:MannelObservablesExp}
    \mathcal{S}_{\rm exp}= -0.04\pm0.06, \quad \mathcal{D}_{\rm exp} = -0.05\pm0.06\:.
\end{align}   
Employing Eqs.~\eqref{eq:AdirBdksphi} and Eq.~\eqref{eq:hadrparRanges22}, we obtain for both $b|_{\rm incl}$ and $b|_{\rm excl}$ the SM predictions
\begin{align}
\label{eq:MannelObservables}
    \mathcal{S}_{\rm SM}= -0.014\pm0.005, \quad \mathcal{D}_{\rm SM} =0 \pm 0.003 \:.
\end{align} 
We observe that the SM predictions and experimental results agree and have similar uncertainties.
Given these results, exploring how potential NP contributions could modify these observables is crucial, which we will discuss in the following section. 

\section{\texorpdfstring{\boldmath New Physics analysis}{}}
\label{sec:NewPhysics}
At the moment, the $B \to \phi K$ data are consistent with our SM estimates. However, given the experimental uncertainties, substantial room remains for NP contributions. Pinning down such contributions is a key motivation in exploring decay channels governed by penguin topologies. This section focuses on  NP contributions entering through $b\to s q\bar{q}$ quark-level processes.

\subsection{Isospin-dependent New-Physics observables}
\label{sec:mannelFleischer}
We start by considering possible isospin-dependent NP contributions to the $\bdphiks$ and $\bplusphikplus$ decays by using the following amplitude parametrization: 
\begin{align}
\label{eq:AmpNPfirst}
A(B^{0,+}\to\phi K^{0,+})  &= \mathcal{P}_{\rm SM} + \sum_i \mathcal{P}^{I=0}_{{\rm NP},i}+ \sum_j \mathcal{P}^{I=1}_{{\rm NP},j} \nonumber \\&= \frac{{\mathcal{A}_{P\pm}^\prime}}{\sqrt{\epsilon}}\left[1+\sum_iv_0^i e^{i\Phi_0^i}e^{i\Delta_0^i}\pm\sum_jv_1^je^{i\Phi_1^j}e^{i\delta_1^i}\right]
\end{align} 
where $\mathcal{P}_{\rm SM}$ was given in Eq.~\eqref{eq:decayamp}. As for the SM decomposition, the subscripts on the right-hand side signify the $\bar{u}u\pm\bar{d}d$ flavour structure and the upper (lower) sign holds for $B_d^0 (B^+)$. We allow for several isospin-dependent NP contributions denoted by $v_{0,1}$ for $I=0,1$ NP magnitudes, $\Phi_{0,1}$ as the CP-violating phase differences, and $\Delta_{0,1}$ as the CP-conserving phase differences. For convenience, we do not show the doubly Cabibbo-suppressed SM hadronic penguin parameter $b$ or its phase $\theta$ here. However, we account for these effects in our numerical analysis using the SM estimates from Eq.~\eqref{eq:hadrparRanges22}.

Expressions including several isospin-dependent NP contributions are rather complicated. Therefore, we focus on the case where the different isospin NP contributions involve the same weak or strong phases. In that case, the amplitude in Eq.~\eqref{eq:AmpNPfirst} simplifies to 
\begin{align}
\label{eq:AmpNP}
A(B^{0,+}\to\phi K^{0,+})  &=\frac{\mathcal{A}_{P\pm}^\prime}{\sqrt{\epsilon}}\left[1+ 
     v_0 e^{i\Delta_0} e^{i\Phi_0} \pm v_1 e^{i\Delta_1} e^{i\Phi_1}\right], 
\end{align} 
where the upper (lower) sign holds for $B_d^0 (B^+)$. 

Using this parameterization, the branching ratio expressions can be written in a form similar to Eq.~\eqref{eq:BRstandardModel} as
\begin{equation}
 \label{eq:BrCn0and+}
    \mathcal{B}(B^{0,+}\to \phi K^{0,+})\equiv \frac{\hat{\mathcal{C}}^{0,+}}{\epsilon} n_{0,+},
\end{equation}
where $\hat{\mathcal{C}}^{0,+}$ is given in Eq.~\eqref{eq:BSM}, and
\begin{align}
\begin{split}
        n_{0,+}&\equiv1 +2 v_0 \cos\Delta_0\cos\Phi_0 \pm 2 v_1 \cos\Delta_1\cos\Phi_1 \\&\quad \pm  2v_0v_1\cos(\Delta_0-\Delta_1)\cos(\Phi_0-\Phi_1) +v_0^2 + v_1^2\:.
\end{split}
\end{align}
The parameter $\mathcal{Z}$, introduced in Eq.~\eqref{eq:ZobsUW}, now becomes
\begin{align}
\label{eq:zmannel}
    \mathcal{Z}&\equiv\frac{n_0-n_+}{n_0+n_+}=2v_1\left[\frac{\cos\Delta_1\cos\Phi_1+v_0\cos\left(\Delta_0-\Delta_1\right)\cos\left(\Phi_0-\Phi_1\right)}{1+2v_0\cos\Delta_0\cos\Phi_0+v_0^2}\right].
\end{align}
Since the SM prediction for $\mathcal{Z}_{\rm SM} $ in Eq.~\eqref{eq:ZSM} is tiny, a non-zero measurement of $\mathcal{Z}$ would indicate $I=1$ NP contributions.

The CP-violating observables $\mathcal{D}$ and $\mathcal{S}$ defined in Eq.~\eqref{eq:SandD} become
\begin{equation}\label{eq:MannelDefinitionsZ}
 %\mathcal{Z} = v_1\frac{f}{e} \ , \quad \quad 
  \mathcal{D}=-2v_1\left[\frac{de-cf}{e^2-f^2v_1^2}\right],  \quad\quad
 \mathcal{S}=-2\left[\frac{ce-df v_1^2}{e^2-f^2v_1^2}\right] ,
\end{equation}
where 
\begin{align}
    c&\equiv v_0\sin\Delta_0\sin\Phi_0\:, \nonumber \\
    d&\equiv \sin\Delta_1\sin\Phi_1+v_0\sin\left(\Delta_0-\Delta_1\right)\cos\left(\Phi_0-\Phi_1\right),\nonumber\\
    e&\equiv 1+2v_0\cos\Delta_0\cos\Phi_0+v_0^2+v_1^2\:,\nonumber\\
    f&\equiv 2\left[\cos\Delta_1\cos\Phi_1+v_0\cos\left(\Delta_0-\Delta_1\right)\cos\left(\Phi_0-\Phi_1\right)\right].
\end{align}
Finally, Eq.~\eqref{eq:UtimesoneplusW} now reads
\begin{align}
   \mathcal{U}\left(1+\mathcal{W}\right) = \frac{n_+}{n_0},
\end{align}
which equals unity in the isospin limit.

\subsection{Expected hierarchies of isospin-dependent New Physics}
Before constraining the NP parameter space, we first consider the generic structure of possible isospin-dependent NP terms \cite{Fleischer:2001pc}. For $I=1$, we expect an $\mathcal{O}(\bar\lambda)$ suppression due to their flavour structure. We note that a similar dynamical suppression also appears in the SM parameters, see Eq.~\eqref{eq:1over0}, but there an additional suppression from the EW penguin structure enters. Compared to the SM $I=1$ terms, the NP effects may thus still be enhanced. For the $I=0$ NP terms, no suppression is expected. Following Ref.~\cite{Fleischer:2001pc}, we then write
\begin{align}
\label{eq:estold}
A(B\to\phi K)  &=\frac{\mathcal{A}_{P\pm}^\prime}{\sqrt{\epsilon}}\left[1+ \mathcal{O}(1)|_{I=0} +   \mathcal{O}(\bar\lambda)|_{I=1} + \mathcal{O}(\epsilon) \right], 
\end{align} 
which is an ``optimistic'' scenario in the sense that NP effects could be large. However, such large NP effects are not supported by the current data. In fact, the $\Delta\mathcal{B}$ values in Eq.~\eqref{eq:DelBplus} and \eqref{eq:DelBKSexcl} leave room for NP effects of $\mathcal{O}(\bar\lambda)$. In addition, comparing again to the $B_d^0\to J/\psi K_S$ modes through \eqref{eq:Deltapsiphidef}, the NP contributions arising at the decay amplitude level in the $B_d^0\to \phi K_{\rm S}$ mode described in Eq.~\eqref{eq:estold} would yield
\begin{equation}
    \Delta \mathcal{A}_{\psi \phi}^{\rm mix} = \mathcal{O}(1)|_{I=0} +  \mathcal{O}(\bar\lambda)|_{I=1} + \ \mathcal{O}(\epsilon)|_{\rm SM} \ .
\end{equation}
Comparing with the experimental data in Eq.~\eqref{eq:psiphiexp}, we note that effects of $\mathcal{O}(\bar\lambda)$ are still allowed by the current data while $\mathcal{O}(1)$ effects were not found. More precise measurements of the mixing-induced CP asymmetry will help to pin down these contributions even further. 

In conclusion, the current data suggest a different hierarchy between the isospin NP components. Keeping the dynamical suppression of the $I=1$ terms in mind\footnote{As discussed before, rescattering effects could enhance the dynamical suppression of the $I=1$ terms. However, there is currently no sign of such large effects in $B$ decays.}, we consider the following scenario \cite{Fleischer:2001pc}:
\begin{align}
\label{eq:estnew}
A(B\to\phi K)  &=\frac{\mathcal{A}_{P\pm}^\prime}{\sqrt{\epsilon}}\left[1+ \mathcal{O}(\bar\lambda)|_{I=0} +   \mathcal{O}(\bar\lambda^2)|_{I=1} + \mathcal{O}(\epsilon) \right], 
\end{align} 
where $\bar\lambda^2 \sim \epsilon$ such that the $I=1$ NP amplitudes would enter at the same level as the doubly Cabibbo-suppressed SM penguin parameters. In that case, it would be challenging to distinguish the $I=1$ NP contributions from SM effects. 

From these arguments, we expect that the isospin observables defined in Eqs.~\eqref{eq:ZobsUW} and Eq.~\eqref{eq:SandD} scale like
\begin{equation}
    \mathcal{S} = \mathcal{O}(\bar\lambda)|_{I=0} + \mathcal{O}(\epsilon)  \ , \quad\mathcal{D} = \mathcal{O}(\bar\lambda^2)|_{I=1} + \mathcal{O}(\epsilon)  \ , \quad\mathcal{Z} =\mathcal{O}(\bar\lambda^2)|_{I=1}  + \mathcal{O}(\epsilon)  \ .
\end{equation}
Comparing with the experimental data for $\mathcal{S}$ and $\mathcal{D}$ in \eqref{eq:MannelObservablesExp}, we find that the current data are in agreement with this pattern. At the same time, the current experimental uncertainty on $\mathcal{D}$ is still sizeable, thereby still allowing for $I=1$ effects of $\mathcal{O}(\bar\lambda)$.

We therefore identify $\mathcal{S}$ and $\mathcal{D}$, i.e., the sum and difference of the direct CP asymmetries in $B_d^0\to \phi K_{\rm S}$ and $B^+\to \phi K^+$, as key observables to further study isospin-dependent NP effects. In the following, we discuss the constraints on the $I=0$ and $I=1$ NP contributions arising from the current data.   

\subsection{\texorpdfstring{ \boldmath Probing $I=0$  New Physics}{}}
First, we probe the $I=0$ NP sector by setting $v_1$ to zero. In this scenario, the NP parts of the amplitudes for the $\bdphiks$ and $\bplusphikplus$ decays are equal. Therefore, we have $\mathcal{Z}_{\rm NP}=0$, $\mathcal{D}_{\rm NP}=0$ and
\begin{equation}
    \mathcal{S}= \frac{-2 v_0 \sin\Delta_0 \sin \Phi_0}{1+2v_0 \cos\Delta_0 \cos\Phi_0 + v_0^2} = \mathcal{A}_{\rm dir}(B_d^0\to \phi K_{\rm S})=\mathcal{A}_{\rm dir}(B^+\to \phi K^+) \ .
\end{equation}
As already stressed above, this is in agreement with the data. 
The expression for the mixing-induced CP asymmetry can be obtained from~\eqref{eq:AmixBdksphi} by substituting
\begin{align}
\label{eq:isospinNPsubstitutions}
    \epsilon b\to v_0, \quad \gamma \to \Phi_{0}, \quad \theta \to \Delta_{0}.
\end{align}

\begin{figure}[t!]
    \centering
    \includegraphics[width=0.99\textwidth]{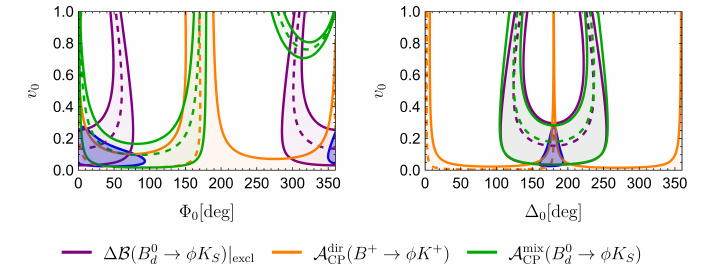}
    \caption{Experimental constraints in a NP scenario with only $I=0$ contributions for $v_{0}$ and the CP-violating phase $\Phi_0$ (left) and CP-conserving phase $\Delta_{0}$ (right). The constraints from $\Delta\mathcal{B}$ (purple), the direct CP asymmetry in $\bplusphikplus$ (orange), and the mixing-induced CP asymmetry in $\bdphiks$ (green) are shown, with the $1\sigma$ best-fit region (blue) obtained from a $\chi^2$ fit.
    }
    \label{fig:I0npPlot}
\end{figure}

The CP asymmetries can then be used to constrain the NP parameters. Here we only use the direct CP asymmetry from the charged mode as its experimental uncertainty is smaller. Adding $B_d^0\to \phi K_{\rm S}$ would not give additional information. In Fig.~\ref{fig:I0npPlot}, we depict the allowed regions in the $\Phi_0$--$v_0$ and $\Delta_0$--$v_0$ planes within $1\sigma$ experimental uncertainties. There, we fix the other phase to its best-fit value, i.e., we fix $\Delta_0$ or $\Phi_0$, respectively. 
These are 
\begin{align}
    v_0=0.15,\quad \Phi_0=27\degree, \quad \Delta_0=174\degree\ .
\end{align}
In addition, the ratio between the NP and SM branching ratio expressions in Eq.~\eqref{eq:DelB} can also be used to further constrain NP effects. The NP effects then enter as
\begin{equation}
\label{eq:fracBRsI=0}
   \Delta \mathcal{B} (B_d^0\to\phi K_{\rm S})=1+2v_0 \cos\Phi_0\cos\Delta_0+v_0^2\ .
\end{equation}
We use $\Delta\mathcal{B}|_{\rm excl}$ from Eq.~\eqref{eq:DelBKSexcl} and assume the SM hadronic penguin parameters in Eq.~\eqref{eq:hadrparRanges22}, noting that --- given they enter in a doubly Cabibbo-suppressed way --- even a sizeable change in these parameters would not significantly alter the shape of the plots. The experimental constraints indicate an allowed range for $v_0$ that is strongly constrained within $1\sigma$ as $v_0 < 0.3$ across the parameter space, indicating possible NP effects at $\mathcal{O}(\bar\lambda)$ following the hierarchy in Eq.~\eqref{eq:estnew}. 

\subsection{\texorpdfstring{ \boldmath Probing $I=1$ New Physics}{}}
\begin{figure}[t!]
    \centering
    \hspace*{0cm}
\includegraphics[width=0.95\textwidth]{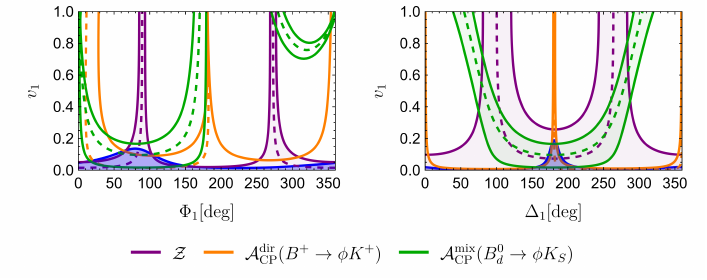}
    \caption{Experimental constraints for a NP scenario with only $I=1$ contributions in the $v_{1}$ --$\Phi_{1}$ plane (left) and the $v_{1}$--$\Delta_{1}$ plane (right). The purple region corresponds to the constraint from $\mathcal{Z}$, while the orange and green regions arise from the direct CP asymmetry in $\bplusphikplus$ and the mixing-induced CP asymmetry in $\bdphiks$, respectively. The blue area indicates the $1\sigma$ best-fit region from a $\chi^2$ fit.}
    \label{fig:I1npPlot}
\end{figure}

Let us now consider a scenario in which NP arises only in the $I=1$ sector, corresponding to $v_0=0$.
In this case, there is a sign difference in the NP contributions to the amplitudes of the $\bdphiks$ and $\bplusphikplus$ modes (see Eq.~\eqref{eq:AmpNP}). As discussed, the $\mathcal{D}$ and $\mathcal{Z}$ isospin observables probe $I=1$ NP directly. Eqs.~\eqref{eq:zmannel}$-$\eqref{eq:MannelDefinitionsZ} then read 
\begin{align}
    \mathcal{Z}_{\rm NP}&=2v_1\cos\Delta_1\cos\Phi_1 + \mathcal{O}(v_1^3)\ ,\\
    \mathcal{D}_{\rm NP}&=-2v_1\sin\Delta_1\sin\Phi_1 + \mathcal{O}(v_1^3)\ ,
    \end{align}
depending linearly on $v_1$. The sum $\mathcal{S}$ of the direct CP asymmetries only depends quadratically on $v_1$:
\begin{align}    
\mathcal{S}_{\rm NP}&=v_1^2\sin2\Delta_1\sin2\Phi_1  + \mathcal{O}(v_1^3)\ .
\end{align}
We can now constrain the NP parameter space using the CP asymmetries and $\mathcal{Z}$. Given the current uncertainties on $\mathcal{S}_{\rm exp}$ in Eq.~\eqref{eq:MannelObservablesExp}, this observable does not constrain the parameter space further. 
In Fig.~\ref{fig:I1npPlot}, we show the available $I=1$ NP parameter space in the $\Phi_1$--$v_1$ and $\Delta_1$--$v_1$ planes given the current experimental constraints. In analogy to Fig.~\ref{fig:I0npPlot}, we fix the other phases to their best-fit values:
\begin{align}
    v_1=0.08,\quad \Phi_1=79\degree,\quad \Delta_1=187\degree\ .
\end{align}
In the plots, we utilized $\mathcal{A}_{\rm CP}^{\rm dir}(B^+\to \phi K^+)$ and note that $\mathcal{D}_{\rm exp}$ would give similar results. We find $v_1<0.2$. In comparison with the $I=0$ case depicted in Fig.~\ref{fig:I0npPlot}, this scenario is constrained more stringently as the SM prediction for the isospin-dependent parameter $\mathcal{Z}$ carries smaller theoretical uncertainties than the branching-ratio difference $\Delta\mathcal{B}$ employed in the $I=0$ scenario. The strong constraint from $\mathcal{Z}$ highlights the sensitivity of this isospin-dependent observable to isospin-dependent NP contributions and motivates further study of this observable. As experimental precision improves, the observables $\mathcal{D}$ and $\mathcal{S}$ will yield valuable insights into the isospin structure of NP effects and the QCD dynamics in the $B \to \phi K$ system. 

\section{Conclusions and Outlook}
\label{sec:conclusions}
 In the SM, $B\to\phi K$ decays originate from penguin loops, rendering them sensitive probes of potential NP contributions. Using a factorization approach, we provided new SM estimates for the branching ratios and CP asymmetries for the neutral $\bdphiks$ and charged $\bplusphikplus$ modes. In addition, we propose the $B_s^0\to \phi K_{\rm S}$ decay to gain further insight into hadronic penguin contributions. At present, no experimental data for this channel are available. By linking this decay to $\bdphiks$, we give SM predictions which can be confronted with future experimental data.
 
Our results are in agreement with the experimental data for the $\bdphiks$ and charged $\bplusphikplus$ modes. In the SM, these modes share the same decay topologies, up to isospin-breaking spectator-quark effects and dynamically suppressed exchange and annihilation topologies. Exploiting the isospin symmetry, we probe isospin-breaking effects in the hadronic parameters through the observable $\mathcal{Z}$. Using current data, we find $\mathcal{Z}=-0.03\pm 0.07$, in agreement with our SM expectation. In addition, we consider the observables $\mathcal{S}$ and $\mathcal{D}$, which probe the sum and difference of the direct CP asymmetries of the $B_d^0\to \phi K^0$ and $B^+\to \phi K^+$ modes, respectively. We find $\mathcal{S}_{\rm exp}=-0.04\pm 0.06$ and $\mathcal{D}_{\rm exp}=-0.05\pm 0.06$, in agreement with our SM predictions, but leaving room for sizeable contributions from physics beyond the SM. Further studying these observables with future data remains of key interest. 

We further studied the constraints on potential isospin-dependent NP contributions. To this end, we employed a model-independent framework to account for such NP effects, decomposed into $I=0$ and $I=1$ components. Based on dynamical arguments, the $I=1$ NP contributions are expected to be suppressed, like their analogous components in the SM. However, since NP contributions may have a very specific isospin structure, we consider scenarios with NP in the $I=0$ and $I=1$ sectors. Using the current experimental data, we find that $I=0$ contributions at the level of $v_0 \leq 0.3$ are still allowed by the current data. For the $I=1$, we find more stringent limits with $v_1 \leq 0.2$. These results provide clear benchmarks for future studies, which allow us to probe the hierarchy of isospin-dependent NP contributions. 

We conclude that the $B\to \phi K$ system offers interesting probes of isospin-dependent NP contributions, especially through the isospin-dependent observables $\mathcal{Z}, \mathcal{S}$ and $\mathcal{D}$. Future measurements at LHCb and Belle~II, and possibly future experiments of the FCC-ee, will improve the precision of CP-violating observables in $B \to \phi K$ processes. Combined with continued theoretical progress, this will enhance the sensitivity to (isospin-dependent) NP effects. %Our isospin-dependent observables will then further sharpen the picture, allowing us to distinguish between $I=0$ and $I=1$ NP contributions.

\section*{Acknowledgements}
This research has been supported by the Netherlands Organization for Scientific Research (NWO).  The work of K.K.V. is supported in part by the Dutch Research Council (NWO) as part of the project Solving Beautiful Puzzles (VI.Vidi.223.083) of the research programme Vidi.

\appendix
\section{Form Factor Calculations}\label{app:ff}
For the form factors entering the $B\to K$ decays discussed in Sec.~\ref{sec:ReviewSM}, we use the Bourrely-Caprini-Lellouch $z$-parameterization \cite{Bourrely:2008za, FlavourLatticeAveragingGroupFLAG:2021npn}. We write
\begin{align}
F_1^{M_iM_f}(q^2)&=\frac{1}{1-q^2/M_\text{pole}^2}\sum_{k=0}^{K-1}b_k^+\left[z(q^2)^k-(-1)^{k-K}\frac{k}{K}z(q^2)^K\right]\ ,
\end{align}
where
\begin{align}
z(q^2)=\frac{\sqrt{t_+-q^2}-\sqrt{t_+-t_0}}{\sqrt{t_+-q^2}+\sqrt{t_+-t_0}}
\end{align}
with \begin{align}t_+&=(M_{M_i}+M_{M_f})^2\ ,\\ t_0&=(M_{M_i}+M_{M_f})(\sqrt{M_{M_i}}-\sqrt{M_{M_f}})^2\ .\end{align} 
For the $B_d^0\to\phi K_S$ decay channel, we have $K=3$ and use $M_i=B_d^0$, $M_f=K^0$, $M_{\rm pole}=5.711$ GeV \cite{Du:2015tda} and $q^2=M_\phi^2$. The parameters $b_i$ are taken from Ref.~\cite{HeavyFlavorAveragingGroupHFLAV:2024ctg} and read as follows: 
\begin{align}
    b_1^+=0.471\pm0.014\quad b_2^+=-0.74 \pm 0.16,\quad b_3^+=0.32\pm0.71,
\end{align}
finding $F_1^{B^0 K^0}(M_\phi^2)=0.391\pm0.021$.

To parameterize the $B_s^0 \to \phi$ form factor, we follow the conventions of Ref.~\cite{Melikhov:2000yu} and write 
\begin{align}
F_1^{B_s^0 \phi}(M_{K_0}^2)&=\frac{f_{A_0}}{\left(1 - \frac{M_{K_0}^2}{M_{B^*_d}^2}\right) \left(1- 
   \frac{M_{K_0}^2 {\sigma_1}_{A_0}}{M_{B^*_d}^2}+\frac{M_{K_0}^4 {\sigma_2}_{A_0}}{M_{B^*_d}^4}\right)} 
   \end{align}
with
\begin{align}
   f_{A_0}=0.42,\quad{\sigma_1}_{A_0}=0.55,\quad {\sigma_2}_{A_0}=0.12,
\end{align}
resulting in $F_1^{B_s^0 \phi}(M_{K_0}^2)=0.43\pm0.04$, where have assumed an error of $10\%$. Note that we have already given this result in Eq.~\eqref{eq:FFBsphi}.

\bibliographystyle{JHEP}
\bibliography{references}

\providecommand{\href}[2]{#2}\begingroup\raggedright\begin{thebibliography}{10}

\bibitem{Fleischer:2024uru}
R.~Fleischer, \emph{{CP violation in B decays: recent developments and future
  perspectives}},
  \href{https://doi.org/10.1140/epjs/s11734-024-01127-0}{\emph{Eur. Phys. J.
  ST} {\bfseries 233} (2024) 391}
  [\href{https://arxiv.org/abs/2402.00710}{{\ttfamily 2402.00710}}].

\bibitem{Fleischer:1996bv}
R.~Fleischer, \emph{{CP violation and the role of electroweak penguins in
  nonleptonic $B$ decays}},
  \href{https://doi.org/10.1142/S0217751X97001432}{\emph{Int. J. Mod. Phys. A}
  {\bfseries 12} (1997) 2459}
  [\href{https://arxiv.org/abs/hep-ph/9612446}{{\ttfamily hep-ph/9612446}}].

\bibitem{Grossman:1997gr}
Y.~Grossman, G.~Isidori and M.P.~Worah, \emph{{CP asymmetry in $B_d \to \phi
  K_S$: Standard model pollution}},
  \href{https://doi.org/10.1103/PhysRevD.58.057504}{\emph{Phys. Rev. D}
  {\bfseries 58} (1998) 057504}
  [\href{https://arxiv.org/abs/hep-ph/9708305}{{\ttfamily hep-ph/9708305}}].

\bibitem{Grossman:2003qp}
Y.~Grossman, Z.~Ligeti, Y.~Nir and H.~Quinn, \emph{{SU(3) relations and the CP
  asymmetries in B decays to $\eta' K_S, \phi K_S$ and $K^+ K^- K_(S)$}},
  \href{https://doi.org/10.1103/PhysRevD.68.015004}{\emph{Phys. Rev. D}
  {\bfseries 68} (2003) 015004}
  [\href{https://arxiv.org/abs/hep-ph/0303171}{{\ttfamily hep-ph/0303171}}].

\bibitem{Fleischer:2001pc}
R.~Fleischer and T.~Mannel, \emph{{Exploring new physics in the $B\to \phi K$
  system}}, \href{https://doi.org/10.1016/S0370-2693(01)00648-7}{\emph{Phys.
  Lett. B} {\bfseries 511} (2001) 240}
  [\href{https://arxiv.org/abs/hep-ph/0103121}{{\ttfamily hep-ph/0103121}}].

\bibitem{Biswas:2024bhn}
A.~Biswas, S.~Descotes-Genon, J.~Matias and G.~Tetlalmatzi-Xolocotzi,
  \emph{{Optimised observables and new physics prospects in the
  penguin-mediated decays
  B$_{d(s)}${\,}{\textrightarrow}{\,}K$^{(*)0}${\ensuremath{\phi}}}},
  \href{https://doi.org/10.1007/JHEP08(2024)030}{\emph{JHEP} {\bfseries 08}
  (2024) 030} [\href{https://arxiv.org/abs/2404.01186}{{\ttfamily
  2404.01186}}].

\bibitem{DeBruyn:2025rhk}
K.~De~Bruyn, R.~Fleischer and E.~Malami, \emph{{How to tame penguins: advancing
  to high-precision measurements of $\phi _d$ and $\phi _s$}},
  \href{https://doi.org/10.1140/epjc/s10052-026-15401-z}{\emph{Eur. Phys. J. C}
  {\bfseries 86} (2026) 215}
  [\href{https://arxiv.org/abs/2505.06102}{{\ttfamily 2505.06102}}].

\bibitem{Faller:2008gt}
S.~Faller, R.~Fleischer and T.~Mannel, \emph{{Precision Physics with $B^0_s \to
  J/\psi \phi$ at the LHC: The Quest for New Physics}},
  \href{https://doi.org/10.1103/PhysRevD.79.014005}{\emph{Phys. Rev. D}
  {\bfseries 79} (2009) 014005}
  [\href{https://arxiv.org/abs/0810.4248}{{\ttfamily 0810.4248}}].

\bibitem{DeBruyn:2022zhw}
K.~De~Bruyn, R.~Fleischer, E.~Malami and P.~van Vliet, \emph{{New physics in
  $B_{q}^0$--$\bar{B}_q^{0}$ mixing: present challenges, prospects, and
  implications for}}, \href{https://doi.org/10.1088/1361-6471/acab1d}{\emph{J.
  Phys. G} {\bfseries 50} (2023) 045003}
  [\href{https://arxiv.org/abs/2208.14910}{{\ttfamily 2208.14910}}].

\bibitem{LHCb:2021dcr}
{\scshape LHCb} collaboration, \emph{{Simultaneous determination of CKM angle
  $\gamma$ and charm mixing parameters}},
  \href{https://doi.org/10.1007/JHEP12(2021)141}{\emph{JHEP} {\bfseries 12}
  (2021) 141} [\href{https://arxiv.org/abs/2110.02350}{{\ttfamily
  2110.02350}}].

\bibitem{Gambino:2019sif}
P.~Gambino, M.~Jung and S.~Schacht, \emph{{The $V_{cb}$ puzzle: An update}},
  \href{https://doi.org/10.1016/j.physletb.2019.06.039}{\emph{Phys. Lett. B}
  {\bfseries 795} (2019) 386}
  [\href{https://arxiv.org/abs/1905.08209}{{\ttfamily 1905.08209}}].

\bibitem{Gambino:2020jvv}
P.~Gambino et~al., \emph{{Challenges in semileptonic $B$ decays}},
  \href{https://doi.org/10.1140/epjc/s10052-020-08490-x}{\emph{Eur. Phys. J. C}
  {\bfseries 80} (2020) 966}
  [\href{https://arxiv.org/abs/2006.07287}{{\ttfamily 2006.07287}}].

\bibitem{Bordone:2019vic}
M.~Bordone, M.~Jung and D.~van Dyk, \emph{{Theory determination of $\bar{B}\to
  D^{(*)}\ell^-\bar\nu$ form factors at $\mathcal{O}(1/m_c^2)$}},
  \href{https://doi.org/10.1140/epjc/s10052-020-7616-4}{\emph{Eur. Phys. J. C}
  {\bfseries 80} (2020) 74} [\href{https://arxiv.org/abs/1908.09398}{{\ttfamily
  1908.09398}}].

\bibitem{Bordone:2021oof}
M.~Bordone, B.~Capdevila and P.~Gambino, \emph{{Three loop calculations and
  inclusive Vcb}},
  \href{https://doi.org/10.1016/j.physletb.2021.136679}{\emph{Phys. Lett. B}
  {\bfseries 822} (2021) 136679}
  [\href{https://arxiv.org/abs/2107.00604}{{\ttfamily 2107.00604}}].

\bibitem{Finauri:2023kte}
G.~Finauri and P.~Gambino, \emph{{The q$^{2}$ moments in inclusive semileptonic
  B decays}}, \href{https://doi.org/10.1007/JHEP02(2024)206}{\emph{JHEP}
  {\bfseries 02} (2024) 206}
  [\href{https://arxiv.org/abs/2310.20324}{{\ttfamily 2310.20324}}].

\bibitem{HeavyFlavorAveragingGroupHFLAV:2024ctg}
{\scshape Heavy Flavor Averaging Group (HFLAV)} collaboration, \emph{{Averages
  of b-hadron, c-hadron, and {\ensuremath{\tau}}-lepton properties as of
  2023}}, \href{https://doi.org/10.1103/x87q-tld5}{\emph{Phys. Rev. D}
  {\bfseries 113} (2026) 012008}
  [\href{https://arxiv.org/abs/2411.18639}{{\ttfamily 2411.18639}}].

\bibitem{Fleischer:2002ys}
R.~Fleischer, \emph{{CP violation in the B system and relations to $K\to \pi
  \nu \bar{\nu}$ decays}},
  \href{https://doi.org/10.1016/S0370-1573(02)00274-0}{\emph{Phys. Rept.}
  {\bfseries 370} (2002) 537}
  [\href{https://arxiv.org/abs/hep-ph/0207108}{{\ttfamily hep-ph/0207108}}].

\bibitem{ParticleDataGroup:2024cfk}
{\scshape Particle Data Group} collaboration, \emph{{Review of particle
  physics}}, \href{https://doi.org/10.1103/PhysRevD.110.030001}{\emph{Phys.
  Rev. D} {\bfseries 110} (2024) 030001}.

\bibitem{Fleischer:1999zi}
R.~Fleischer, \emph{{Extracting CKM phases from angular distributions of B(d,s)
  decays into admixtures of CP eigenstates}},
  \href{https://doi.org/10.1103/PhysRevD.60.073008}{\emph{Phys. Rev. D}
  {\bfseries 60} (1999) 073008}
  [\href{https://arxiv.org/abs/hep-ph/9903540}{{\ttfamily hep-ph/9903540}}].

\bibitem{Barel:2020jvf}
M.Z.~Barel, K.~De~Bruyn, R.~Fleischer and E.~Malami, \emph{{In pursuit of new
  physics with $B_d^0\to J/\psi K^0$ and $B_s^0\to J/\psi\phi$ decays at the
  high-precision Frontier}},
  \href{https://doi.org/10.1088/1361-6471/abf2a2}{\emph{J. Phys. G} {\bfseries
  48} (2021) 065002} [\href{https://arxiv.org/abs/2010.14423}{{\ttfamily
  2010.14423}}].

\bibitem{Barel:2022wfr}
M.Z.~Barel, K.~De~Bruyn, R.~Fleischer and E.~Malami, \emph{{Penguin Effects in
  $B_d^0\to J/\psi K^0_S$ and $B_s^0\to J/\psi\phi$}},  in \emph{{11th
  International Workshop on the CKM Unitarity Triangle}}, 3, 2022
  [\href{https://arxiv.org/abs/2203.14652}{{\ttfamily 2203.14652}}].

\bibitem{Fleischer:1992gp}
R.~Fleischer, \emph{{CP violating asymmetries in penguin induced B meson decays
  beyond the leading logarithmic approximation}},
  \href{https://doi.org/10.1007/BF01557708}{\emph{Z. Phys. C} {\bfseries 58}
  (1993) 483}.

\bibitem{Fleischer:1993gr}
R.~Fleischer, \emph{{Electroweak Penguin effects beyond leading logarithms in
  the B meson decays $B\to K^-\phi$ and $B^-\to \pi^- \bar{K}^0$}},
  \href{https://doi.org/10.1007/BF01559527}{\emph{Z. Phys. C} {\bfseries 62}
  (1994) 81}.

\bibitem{Buras:1998raa}
A.J.~Buras, \emph{{Weak Hamiltonian, CP violation and rare decays}},  in
  \emph{{Les Houches Summer School in Theoretical Physics, Session 68: Probing
  the Standard Model of Particle Interactions}}, pp.~281--539, 6, 1998
  [\href{https://arxiv.org/abs/hep-ph/9806471}{{\ttfamily hep-ph/9806471}}].

\bibitem{Bander:1979px}
M.~Bander, D.~Silverman and A.~Soni, \emph{{CP Noninvariance in the Decays of
  Heavy Charged Quark Systems}},
  \href{https://doi.org/10.1103/PhysRevLett.43.242}{\emph{Phys. Rev. Lett.}
  {\bfseries 43} (1979) 242}.

\bibitem{Gerard:1988jj}
J.-M.~Gerard and W.-S.~Hou, \emph{{CP Nonconservation and CPT: A Reassessment
  of Loop Effects in Charmless B Decays}},
  \href{https://doi.org/10.1103/PhysRevLett.62.855}{\emph{Phys. Rev. Lett.}
  {\bfseries 62} (1989) 855}.

\bibitem{Isgur:1990kf}
N.~Isgur and M.B.~Wise, \emph{{Relationship Between Form-factors in
  Semileptonic $\bar{B}$ and $D$ Decays and Exclusive Rare $\bar{B}$ Meson
  Decays}}, \href{https://doi.org/10.1103/PhysRevD.42.2388}{\emph{Phys. Rev. D}
  {\bfseries 42} (1990) 2388}.

\bibitem{FlavourLatticeAveragingGroupFLAG:2021npn}
{\scshape Flavour Lattice Averaging Group (FLAG)} collaboration, \emph{{FLAG
  Review 2021}},
  \href{https://doi.org/10.1140/epjc/s10052-022-10536-1}{\emph{Eur. Phys. J. C}
  {\bfseries 82} (2022) 869}
  [\href{https://arxiv.org/abs/2111.09849}{{\ttfamily 2111.09849}}].

\bibitem{Melikhov:2000yu}
D.~Melikhov and B.~Stech, \emph{{Weak form-factors for heavy meson decays: An
  Update}}, \href{https://doi.org/10.1103/PhysRevD.62.014006}{\emph{Phys. Rev.
  D} {\bfseries 62} (2000) 014006}
  [\href{https://arxiv.org/abs/hep-ph/0001113}{{\ttfamily hep-ph/0001113}}].

\bibitem{DeBruyn:2012wj}
K.~De~Bruyn, R.~Fleischer, R.~Knegjens, P.~Koppenburg, M.~Merk and N.~Tuning,
  \emph{{Branching Ratio Measurements of $B_s$ Decays}},
  \href{https://doi.org/10.1103/PhysRevD.86.014027}{\emph{Phys. Rev. D}
  {\bfseries 86} (2012) 014027}
  [\href{https://arxiv.org/abs/1204.1735}{{\ttfamily 1204.1735}}].

\bibitem{DeBruyn:2014oga}
K.~De~Bruyn and R.~Fleischer, \emph{{A Roadmap to Control Penguin Effects in
  $B^0_d\to J/\psi K_{\rm S}^0$ and $B^0_s\to J/\psi \phi$}},
  \href{https://doi.org/10.1007/JHEP03(2015)145}{\emph{JHEP} {\bfseries 03}
  (2015) 145} [\href{https://arxiv.org/abs/1412.6834}{{\ttfamily 1412.6834}}].

\bibitem{Bourrely:2008za}
C.~Bourrely, I.~Caprini and L.~Lellouch, \emph{{Model-independent description
  of $B\to \pi \ell \nu$ decays and a determination of $|V_{ub}|$}},
  \href{https://doi.org/10.1103/PhysRevD.82.099902}{\emph{Phys. Rev. D}
  {\bfseries 79} (2009) 013008}
  [\href{https://arxiv.org/abs/0807.2722}{{\ttfamily 0807.2722}}].

\bibitem{Fleischer:2001cw}
R.~Fleischer and T.~Mannel, \emph{{General analysis of new physics in $B \to
  J/\psi K$}}, \href{https://doi.org/10.1016/S0370-2693(01)00346-X}{\emph{Phys.
  Lett. B} {\bfseries 506} (2001) 311}
  [\href{https://arxiv.org/abs/hep-ph/0101276}{{\ttfamily hep-ph/0101276}}].

\bibitem{Gronau:1995hn}
M.~Gronau, O.F.~Hernandez, D.~London and J.L.~Rosner, \emph{{Electroweak
  penguins and two-body B decays}},
  \href{https://doi.org/10.1103/PhysRevD.52.6374}{\emph{Phys. Rev. D}
  {\bfseries 52} (1995) 6374}
  [\href{https://arxiv.org/abs/hep-ph/9504327}{{\ttfamily hep-ph/9504327}}].

\bibitem{Buras:1995pz}
A.J.~Buras and R.~Fleischer, \emph{{Towards the control over electroweak
  penguins in nonleptonic B decays}},
  \href{https://doi.org/10.1016/0370-2693(95)01279-6}{\emph{Phys. Lett. B}
  {\bfseries 365} (1996) 390}
  [\href{https://arxiv.org/abs/hep-ph/9507303}{{\ttfamily hep-ph/9507303}}].

\bibitem{Du:2015tda}
D.~Du, A.X.~El-Khadra, S.~Gottlieb, A.S.~Kronfeld, J.~Laiho, E.~Lunghi et~al.,
  \emph{{Phenomenology of semileptonic B-meson decays with form factors from
  lattice QCD}}, \href{https://doi.org/10.1103/PhysRevD.93.034005}{\emph{Phys.
  Rev. D} {\bfseries 93} (2016) 034005}
  [\href{https://arxiv.org/abs/1510.02349}{{\ttfamily 1510.02349}}].

\end{thebibliography}\endgroup

\end{document}